\documentclass[fleqn,usenatbib]{mnras}

\usepackage{newtxtext,newtxmath}
\usepackage[T1]{fontenc}

\DeclareRobustCommand{\VAN}[3]{#2}
\let\VANthebibliography\thebibliography
\def\thebibliography{\DeclareRobustCommand{\VAN}[3]{##3}\VANthebibliography}

\usepackage{graphicx} 
\usepackage{amsmath} 

\usepackage{amssymb} 
\usepackage{xcolor}  
\usepackage{lipsum}  
\usepackage{caption} 
\usepackage{siunitx} 
\usepackage{courier} 
\usepackage{longtable} 
\usepackage{makecell} 
\usepackage{ragged2e} 
\usepackage{gensymb} 
\usepackage[title]{appendix}

\newcommand{\vdag}{^\dagger}            
\newcommand{\vddag}{^\ddagger}            

\newcommand{\Teffprim} {$T_{\mathrm{eff},\mathrm{A}}$} 
\newcommand{\loggprim} {log\,$g_{\mathrm{A}}$} 
\newcommand{\fehprim} {[Fe/H]$_{\mathrm{A}}$} 
\newcommand{\MHprim} {[M/H]$_{\mathrm{A}}$} 
\newcommand{\Msun}{M$_\mathrm{\odot}$}   
\newcommand{\Rsun}{R$_\mathrm{\odot}$}   
\newcommand{\uni}{$\mathcal{U}$}        
\newcommand{\norm}{$\mathcal{N}$}        
\newcommand{\kms}{$\mathrm{~km\,s^{-1}}$}  
\newcommand{\ms}{$\mathrm{~m\,s^{-1}}$}    
\newcommand{\masyr}{$\mathrm{~mas\,yr^{-1}}$}    
\newcommand{\cgs}{$\mathrm{~g\,cm^{-3}}$}    
\newcommand{\BJDTDB}{BJD$_\mathrm{TDB}$} 
\newcommand{\vsini}{$\upsilon\,\mathrm{sin}\,i\,$} 
\newcommand{\veqprim}{$\upsilon_A$} 
\newcommand{\vsiniprim}{$\upsilon_A\,\mathrm{sin}\,i\,$} 
\newcommand{\ispec}{\textbf{\texttt{iSpec}}}      
\newcommand{\radvel}{\textbf{\texttt{RADVEL}}}      
\newcommand{\exofast}{\textbf{\texttt{EXOFAST}}}      
\newcommand{\ariadne}{\textbf{\texttt{Ariadne}}}      
\newcommand{\allesfitter}{\textbf{\texttt{Allesfitter}}}      
\newcommand{\Lightkurve}{\textbf{\texttt{Lightkurve}}}      
\newcommand{\dynesty}{\textbf{\texttt{Dynesty}}}      
\newcommand{\minaus}{{\textsc{Minerva}}}

\title[TIC 48227288, TIC 339607421 obliquity]{The Spin-orbit alignment of two short period eclipsing binary systems}

\author[T. Wells et al.]{Tony Wells,$^{1}$\thanks{E-mail: tony.wells@usq.edu.au}
B. C. Addison,$^{1,2}$
R.A. Wittenmyer,$^{1}$
Duncan J. Wright,$^{1}$
Tyler R. Fairnington,$^{1}$
\newauthor
Jason Dittmann,$^{3}$
Jonathan Horner,$^{1}$
Stephen R. Kane,$^{4}$
John Kielkopf,$^{5}$
Peter Plavchan,$^{6}$
Avi Shporer,$^{7}$
\\
$^{1}$University of Southern Queensland, Centre for Astrophysics, West St, Toowoomba, QLD 4350, Australia\\
$^{2}$Swinburne University of Technology, Centre for Astrophysics and Supercomputing, John Street, Hawthorn, VIC 3122, Australia\\
$^{3}$Department of Astronomy, University of Florida, 211 Bryant Space Science Center, Gainesville, FL, 32611, USA\\
$^{4}$Department of Earth and Planetary Sciences, University of California, Riverside, CA 92521, USA\\
$^{5}$Department of Physics and Astronomy, University of Louisville, Louisville, KY 40292, USA\\
$^{6}$George Mason University, 4400 University Drive MS 3F3, Fairfax, VA 22030, USA\\
$^{7}$Department of Physics and Kavli Institute for Astrophysics and Space Research, Massachusetts Institute of Technology, Cambridge, MA 02139, USA\\
}

\date{Accepted XXX. Received YYY; in original form ZZZ}

\pubyear{2024}

\begin{document}
\label{firstpage}
\pagerange{\pageref{firstpage}--\pageref{lastpage}}
\maketitle


\begin{abstract}

We present a joint analysis of TESS photometry and radial velocity measurements obtained from the Minerva-Australis facility for two short-period eclipsing binaries, TIC 48227288 and TIC 339607421. TIC 339607421 hosts an M-dwarf companion ($M_B = 0.294 \pm 0.013$ \Msun{}, $R_B = 0.291 \pm 0.006$ \Rsun{}) orbiting an F6V star ($M_A=1.09 \pm 0.04$ \Msun{}, $R_A=1.21^{+0.03}_{-0.02}$ \Rsun{}). While TIC 48227288 contains a late K class companion ($M_B=0.635 \pm 0.037$ \Msun{}, $R_B = 0.605 \pm 0.011$ \Rsun{}) orbiting an F3V star ($M_A = 1.36^{+0.06}_{-0.08}$ \Msun{}, $R_A = 1.61 \pm 0.03$ \Rsun{}). Both companions follow short period, near-circular orbits ($P_B$ = 2.4–3.0 d, $e \approx 0.001$). Sky-projected obliquities for each system were derived using a classical analysis of the RV perturbation and the Reloaded Rossiter-McLaughlin (RRM) technique. The classical method indicates minor spin–orbit misalignment for both systems ($\lambda_A = -14.7^{+5.4}_{-5.9} \degree $ and $-17.8^{+1.9}_{-2.0} \degree $ for TIC 339607421 and TIC 48227288, respectively). The RRM analysis yields smaller obliquities ($\lambda_A  = -8.2 \pm 0.2 \degree$ and $-9.5 \pm 0.2 \degree$ respectively), but confirms the minor misalignment inferred from the classical analysis. The findings of misaligned, circular orbits are notable even though the misalignments are not large, and suggest potential gaps in current models of binary formation and orbital evolution. As such, further investigation of these and similar systems appears warranted.

\end{abstract}




\section{Introduction}

The concept of gravitationally bound multiple star systems dates back to the latter half of the 18th century, when John Michell applied statistical methods to argue that double or multiple stars, rather than being mere chance alignments, were much more likely to be in close physical proximity \citep{Michell_1767:1590}. Soon after, William Herschel cataloged over 700 binaries, \citep{Hershel_1782:1589,Herschel_1785:1591}, before proposing that such stars orbit one another \citep{Herschel_1802:1592}. Around the same time John Goodricke, \citep{Goodricke_1783:251,Goodricke_1784:1672}, attributed the  periodic dimming of $\beta$ Persei  to the regular eclipsing of the star by a close, unseen orbiting companion - a foundational insight for the study of eclipsing binaries.  

Multiple star systems are now known to be common: approximately 44\% of all FGK stars, 26\% of M class stars and $\gtrsim$ 60-80\% of early type stars exist in multiple configurations \citep{Duchene_2013:1575,Winters_2019:1676}. High  multiplicity is also observed among pre-main sequence stars \citep[e.g.][]{Mathieu_1994:1576,Simon_1995:1597,Ghez_1997:1598,Tohline_2002:1599}. Despite this prevalence, our understanding of the processes that lead to the formation of such systems is incomplete. Three principal formation pathways have been identified \citep{Tohline_2002:1599,Duchene_2013:1575,Tokovinin_2020:1593,Kuruwita_2023:1578}:
(1) Core fragmentation of turbulent pre-stellar cores during, or shortly after, the free fall collapse phase;  (2) Disk fragmentation via gravitational instability within massive circumstellar disks; and (3) Dynamical capture of single, unbound stars formed in nearby stellar systems. Core fragmentation is believed to produce wide binaries (with separations of $\sim$100 to $\sim$10,000 AU), whereas disk fragmentation results in closer binaries on the scale of the disk radius (<~100 AU) \citep{Kuruwita_2023:1578}. It is widely acknowledged that these initial orbital configurations can subsequently evolve due to interactions with tertiary companions \citep{Kozai_1962:307}, circumstellar disks \citep{Tokovinin_2020:1593}, and the galactic tidal field \citep{Priyatikanto_2016:1600}, or following the ejection of a third body from the system \citep{Armitage_1997:1601}.

This study focuses on two close binary systems with orbital periods of 2-3 days which current theory suggests are likely products of disk fragmentation. At first glance, it would therefore seem likely that the spin and orbital axes of each member of the binary would be aligned as they originate from the same regions of their respective molecular clouds. However, as noted by \citet{Albrecht_2012:1212}, the primary and companion stars would have been significantly larger during the pre-main sequence phase thereby precluding the existence of such compact orbits at that time. The close proximity therefore suggests that significant orbital evolution has taken place post-formation.

Orbital evolution may involve chaotic processes known to produce spin-orbit misalignment, including Kozai-Lidov interactions with a wide orbiting tertiary body \citep{Kozai_1962:307}, disk warping during accretion \citep{Bate_2010:1584} or magnetic interaction between the two stars and the circumstellar disk \citep{Lai_2011:1585}. Alternatively, close binaries may be produced by accretion driven inward migration  following disk fragmentation \citep{Tokovinin_2020:1593}. Measuring spin-orbit alignment (obliquity) in these systems should therefore constrain formation and evolutionary pathways: low obliquities suggest quiescent histories (or later alignment via tidal interactions), whereas high obliquities imply a more chaotic formation history.

Obliquities are now commonly measured in exoplanet systems, \citep[see, for example,][]{Addison_2013:450,Addison_2016:447,Clark_2022:1338},  yet relatively few studies have explored spin-orbit alignment in stellar binaries. This is somewhat surprising given that obliquity is commonly determined by examination of the Rossiter-McLaughlin (RM) effect, which was first observed a century ago in the $\beta$ Lyrae \citep{Rossiter_1924:260} and Algol \citep{McLaughlin_1924:228} systems. The paucity of binary star obliquity studies stems from two main factors: (1) instrument precision has only recently improved to the point that the measurement of spin orbit alignment for more slowly rotating stars is possible \citep{Triaud_2013:434} and,  (2) binary system spectra are often contaminated by light from the companion, complicating interpretation (there is no such issue when observing transiting exoplanets). To mitigate the latter complication, we  follow the strategy of \citet{Triaud_2013:434} and focus on single-lined eclipsing binaries (SB1s). This approach naturally leads to the study of binary star systems with large mass ratios and allows comparison to formation pathways more commonly observed in the establishment of brown dwarf and exoplanetary systems.

The RM effect manifests as distortions in rotationally broadened absorption lines during eclipses, when the transiting companion sequentially obscures blue and red shifted limbs of the host star.  These distortions allow reconstruction of the path of the occulting body. Traditionally, obliquities are inferred from the perturbation in radial velocities arising from the resulting movement in the spectral line centroid \citep{Triaud_2017:94}. More recent approaches, including Doppler tomography \citep{Zhou_2016:379,Zhou_2016:348,Zhou_2017:780} and the Reloaded RM (RRM) \citep{Cegla_2016:368,Bourrier_2020:459,Kunovac_2021:754,Doyle_2023:1467}, directly model distortions. Doppler tomography involves fitting a spectral model to the disk integrated lines to account for the signature of the companion while RRM isolates the stellar line occulted by the companion and directly determines the local RV centroid. Another direct analysis technique, RM revolutions (RMR), has recently been developed \citep{Bourrier_2012:1274} for systems with small companions that are less suitable for RRM analysis.

Table \ref{table A1:previous studies}, (partially reproduced from \citet{Albrecht_2011:1201}), lists obliquity measurements reported for stellar binaries. Early studies are dominated by quantitative (aligned/misaligned) estimates of spin-orbit alignment \citep[see, for example][]{Twigg_1979:1439}. In an early quantitative study \citet{Hale_1994:1258}  examined 73 multiple star systems and concluded that all solar-type binary systems with separations less than 40 AU were generally aligned. However, a re-analysis, \citep{Justesen_2020:1238}, cast doubt on Hale's findings and thus they have been omitted from Table \ref{table A1:previous studies}. More recent surveys include the BANANA ("Binaries Are Not Always Neatly Aligned") project \citep{Albrecht_2007:1216,Albrecht_2009:1200,Albrecht_2011:1201,Albrecht_2013:1215,Albrecht_2014:1214,Winn_2011:1203}  which modeled the RM effect observed for hot binary systems comprising stars of similar masses, and the EBLM (“Eclipsing Binary, Low Mass”) project, \citep{Triaud_2013:434,Kunovac_2020:1209}, where RRM has been employed to derive obliquities for systems containing low mass companion stars. Other methods used to determine binary star obliquity include apsidal motion \citep{Marcussen_2022:1213}, and gravity darkening \citep{Zhou_2013:1239,Ahlers_2014:1251,Liang_2022:1249}. While most results indicate aligned configurations, exceptions exist, including AI Phe \citep{Sybilski_2018:1247}, CV Velorum \citep{Albrecht_2014:1214} and DI Herculis \citep{Albrecht_2009:1200,Philippov_2013:1253,Liang_2022:1249}. More recently, studies of binary star populations with periods ranging from 50 to 3000 days, \citep{Marcussen_2024:1690, Smith_2024:1691}, suggest that the likelihood of misalignment increases with increasing eccentricity for these wide orbiting systems. 

In this work, we expand the current sample of binary system obliquity measurements by analyzing two SB1 systems: TIC 48227288 and TIC 339607421. We derive sky-projected obliquities ($\lambda$) and orbital parameters from joint analyses of photometric data collected by NASA’s \textit{Transiting Exoplanet Survey Satellite} (\textit{TESS}) and ground-based spectroscopic observations carried out at the \minaus{}-Australis facility.  Two approaches are taken to estimate obliquity. Obliquities are firstly obtained through a classical analysis of the RV perturbation. Further obliquity estimates are then derived from a more direct analysis of the spectroscopic data via the RRM technique. The structure of this paper is as follows: Section \ref{sec:Observations} details the observational data and reduction methods; Section \ref{Section: Analysis of data} presents stellar parameters and joint analysis results; Section \ref{Section: Discussion} discusses our findings in context of previous studies; and Section \ref{section:conclusions} summarizes our main conclusions.


\section{Observations}\label{sec:Observations}

Space-based photometric observations of the two target systems were conducted intermittently over several years using \textit{TESS} (Section \ref{subsec:Photometry}) with high resolution spectroscopic observations subsequently carried out at the \minaus{}-Australis observatory (Section~\ref{subsec:Spectroscopy}). Observation details and data processing procedures are presented below.


\subsection{\textit{TESS} Photometry} \label{subsec:Photometry}


\begin{table} 
\small
\renewcommand{\arraystretch}{1.1}
\caption{ Details of \textit{TESS} observations of TIC 48227288 and TIC 339607421. }
\label{tab:TESS details}
\begin{tabular}{lccc}
\hline\hline
\noalign{\vskip 3px}
 & Camera & CCD Chip & Dates \\

\noalign{\vskip 3px}
\hline
\noalign{\vskip 3px}


\textbf{TIC 48227288}  &   &   & \\
Sector 11              & 1 & 3 & April 23 - May 20, 2019   \\
Sector 38              & 1 & 4 & April 29 - May 26, 2021  \\
Sector 65              & 1 & 4 & May 4 - May 30, 2023  \\
\textbf{TIC 339607421} &   &   &   \\
Sector 2               & 3 & 4 & Aug. 23 - Sept. 20, 2018  \\
Sector 4               & 3 & 3 & Oct. 19 - Nov. 14, 2018  \\
Sector 29              & 3 & 4 & Aug. 29 - Sept. 21, 2020  \\
Sector 30              & 3 & 2 & Sept. 23 - Oct. 20, 2020  \\
Sector 69              & 3 & 1 & Aug. 25 - Sept. 20, 2023  \\


\noalign{\vskip 3px}
\hline

\end{tabular}

\end{table}

\begin{figure*}
\resizebox{\hsize}{!} 
{\includegraphics[]{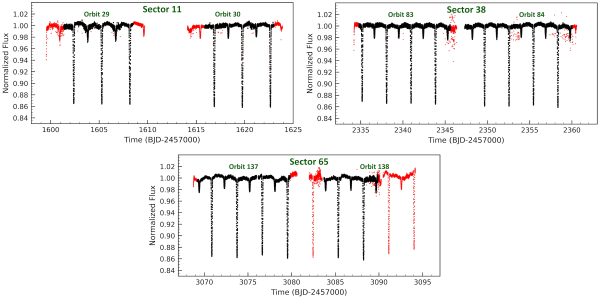}}
\caption{TIC 48227288 \textit{TESS} PDCSAP light curves. Data shown in black were used in the joint \allesfitter{} analysis. Red data were excluded  due to quality issues. The data shown in this figure are available online in machine readable format. }
\label{fig:TIC48227288 LC}
         
\end{figure*}

\begin{figure*}
\resizebox{\hsize}{!}
{\includegraphics[]{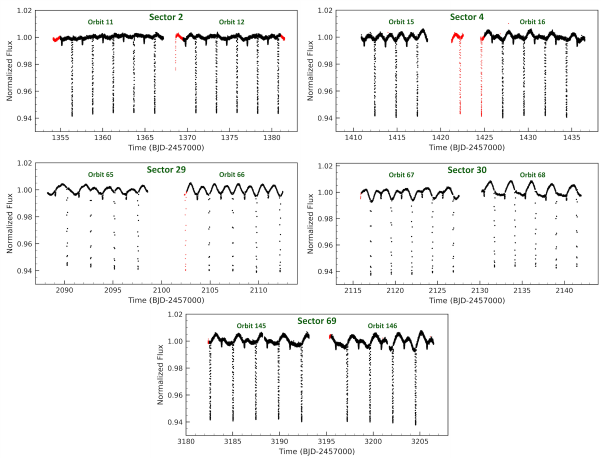}}
        
         \caption{TIC 339607421 \textit{TESS} PDCSAP light curves. Data shown in black were used in the joint \allesfitter{} analysis. Red data were excluded  due to quality issues. The data shown in this figure are available online in machine readable format.}
         
         \label{fig:TIC339607421 LC}
         
\end{figure*}

\textit{TESS} observed TIC 48227288 in sectors 11, 38 and 65 and TIC 339607421 in sectors 2, 4, 29, 30 and 69. Cameras and CCD chips used, along with the corresponding observation windows, are listed in Table \ref{tab:TESS details}. All TIC 48227288 observations were obtained in 120~s cadence mode. A total of 23 primary and 24 secondary eclipses were recorded for TIC 48227288 with depths of $\sim$135 and  $\sim$20 ppt respectively (Figure~\ref{fig:TIC48227288 LC}). TIC 339607421 was observed in 120~s cadence mode in  sectors 2, 4 and 69 and 600 s cadence mode in sectors 29 and 30. A total of 47 primary and 45 secondary eclipses were observed for this target, with depths of approximately 60 and 5 ppt respectively (Figure~\ref{fig:TIC339607421 LC}).

For both systems we used Pre-search Data Conditioning Simple Aperture Photometry (\textit{PDCSAP}) photometric data retrieved using the \Lightkurve{}\footnote{\href{https://github.com/lightkurve/lightkurve}{https://github.com/lightkurve/lightkurve}} package \citep{lightkurve_2018:1568}. Data points with non-zero quality flags were removed, and any partially observed eclipses excluded. Intra-sector gaps in data continuity (e.g. due to momentum dumps) were identified, and light curve segments between such gaps independently normalized using \Lightkurve{} routines (black points in Figures \ref{fig:TIC48227288 LC} and \ref{fig:TIC339607421 LC}).  For modeling purposes, the TIC 48227288 photometric data were treated as six independent data sets, while the TIC 339607421 data were divided into ten independent segments.


\begin{table} 
\small
\renewcommand{\arraystretch}{1.1}
\caption{ TIC 48227288 radial velocities determined from \minaus{} observations. 'RM' appended to an instrument entry indicates that the observation was made during an eclipse.}
\label{tab:TIC88rv}
\begin{tabular}{lccc}
\hline\hline
\noalign{\vskip 3px}
Time    & Velocity & Uncertainty & Instrument \\

[BJD]   & [\ms]    & [\ms]       &            \\
\noalign{\vskip 3px}
\hline
\noalign{\vskip 3px}


2459757.044306 & 46976 & 152 & \minaus T1  \\ 
2459769.010814 & 56657 & 141 & \minaus T1  \\
2459786.95997 & 7544 & 144 & \minaus T1RM  \\
2459795.988807 & -33469 & 147 & \minaus T1  \\
2460008.183256 & -45992 & 206 & \minaus T1  \\
2459757.044306 & 46912 & 151 & \minaus T3  \\
2459769.010814 & 56459 & 154 & \minaus T3  \\
2459786.95997 & 7182 & 148 & \minaus T3RM  \\
2459795.988807 & -33722 & 153 & \minaus T3  \\
2460008.183256 & -46072 & 164 & \minaus T3  \\
2460021.263515 & 40440 & 155 & \minaus T3  \\
2460035.995702 & 12467 & 181 & \minaus T3  \\
2460036.014002 & 10622 & 166 & \minaus T3RM  \\
2460036.032255 & 8517 & 164 & \minaus T3RM  \\
2460036.050509 & 6917 & 157 & \minaus T3RM  \\
2460036.068762 & 5093 & 166 & \minaus T3RM  \\


\noalign{\vskip 3px}
\hline

\end{tabular}
\vskip 3px
Note - The above table is published in full in machine readable form online. The portion shown here is to provide a guide to form and content.
\end{table}



\begin{table} 
\small
\renewcommand{\arraystretch}{1.1}
\caption{ TIC 339607421 radial velocities determined from \minaus{} observations. 'RM' appended to an instrument entry indicates that the observation was made during an eclipse.}
\label{tab:TIC339rv}
\begin{tabular}{lccc}
\hline\hline
\noalign{\vskip 3px}
Time    & Velocity & Uncertainty & Instrument \\

[BJD]   & [\ms]    & [\ms]       &            \\
\noalign{\vskip 3px}
\hline
\noalign{\vskip 3px}


2459845.02718 & 26843 & 237 & \minaus T1  \\
2459916.981703 & 2708 & 288 & \minaus T1  \\
2459922.065061 & -12999 & 253 & \minaus T1  \\
2459928.930804 & 25874 & 224 & \minaus T1  \\
2459933.016964 & 43458 & 238 & \minaus T1  \\
2459941.977936 & -18524 & 257 & \minaus T1  \\
2459952.957495 & 50274 & 233 & \minaus T1  \\
2459980.948251 & -19570 & 253 & \minaus T1  \\
2459922.065061 & -12873 & 237 & \minaus T3  \\
2459933.016964 & 43960 & 213 & \minaus T3  \\
2459952.957495 & 50078 & 216 & \minaus T3  \\
2459980.948251 & -19320 & 247 & \minaus T3  \\
2460021.886871 & -2292 & 240 & \minaus T3  \\
2460136.194679 & 23296 & 251 & \minaus T3  \\
2460136.217365 & 20752 & 258 & \minaus T3RM  \\
2460136.240028 & 19564 & 256 & \minaus T3RM  \\


\noalign{\vskip 3px}
\hline

\end{tabular}
\vskip 3px
Note - The above table is published in full in machine readable form online. The portion shown here is to provide a guide to form and content.
\end{table}


\subsection{\minaus{} spectroscopy} 

\label{subsec:Spectroscopy}
High resolution spectroscopic observations of TIC 48227288 and TIC 339607421 were conducted at the \minaus{}-Australis facility, located at the Mt Kent Observatory in Queensland, Australia \citep{Addison_2019:91,Wittenmyer_2018:278}. The facility comprises an array of remotely accessible, independent 0.7~m PlaneWave CDK700 telescopes. Four telescopes, (T1, T3, T4 and T5), simultaneously feed starlight via fiber optics to a shared KiwiSpec R4-100 high resolution spectrograph, operating at a resolution of R$\approx$80,000 over a wavelength range of 480 to 630~nm. 

TIC 48227288 was observed across 129 epochs between June 26, 2022 and August 31, 2023, including coverage of primary eclipses that took place on 1st April 2023, (10 observations on T3 and 6 on T4), and 30th April 2023, (9 observations on T3 and 10 on T4). 

TIC 339607421 was observed over 144 epochs from September 22, 2022 to September 9, 2023, including coverage of primary eclipses on July 10 2023, (10 observations on T3 and 9 each on T4 and T5), August 23 2023, (6 observations on T3 and 7 on T4), and 9 September 2023, (10 observations each on T3 and T4).  

Two simultaneous calibration fibers, illuminated by a quartz lamp shining through an iodine cell, were used to monitor and correct instrumental drifts.  Radial velocities (RVs) were extracted by cross-correlating each spectrum with a synthetic mask chosen to closely match the primary stellar type as determined via \ispec{}/\ariadne{} analysis (see Section \ref{subsection:host star analyses}). An F3V mask was used in the analysis of TIC 48227288 spectra and an F6V mask for TIC 339607421. Final RV measurements are provided in Tables \ref{tab:TIC88rv} and \ref{tab:TIC339rv}. In this study, radial velocities originating from the individual \minaus{} telescopes are modeled on the assumption that they originate from independent instruments.


\section{Analysis} \label{Section: Analysis of data}


\subsection{Primary star characterisation} \label{subsection:host star analyses}

Before companion stars properties (hereafter denoted with subscript 'B') can be accurately determined, primary star parameters (designated with subscript 'A') must be accurately determined. The following procedure was adopted to determine the primary star parameters.

To begin the primary star characterization process, the 10 highest signal-to-noise ratio (SNR) \minaus{} spectra  were selected for for each target to be analyzed using the \ispec{} software package\footnote{\href{https://github.com/marblestation/iSpec}{https://github.com/marblestation/iSpec}}  \citep{blanco_2014:1396,Blanco_2019:1397}. Spectra were first set in the rest frame before blaze corrections to individual orders were performed using flat field observations carried out previously at \minaus{}-Australis. Sections of the spectra likely to be affected by telluric contamination were then removed using the \ispec{} telluric filter with margins set to the default value of $\pm 30 $~\kms{}. Subsequent normalization of the spectra was complicated by the presence of a pronounced H$\beta$ absorption feature ($\sim$484 to 488~nm) that dominated the first order of the \minaus{} spectra  and hindered spline-based continuum fitting. To address this, a two step normalization process was adopted: First, an approximate normalization of the continuum was performed using the \ispec{} spline function and initial estimates of the primary star's effective temperature, \Teffprim{}, overall metallicity, \MHprim{}, surface gravity, \loggprim{}, and projected rate of rotation, \vsiniprim{} determined using \ispec{'s} grid-modeling routine. A synthetic spectral analysis was then performed to obtain an initial estimate the metallicity, \fehprim{}. 

\begin{figure}
\centering
\includegraphics[width=1\columnwidth]{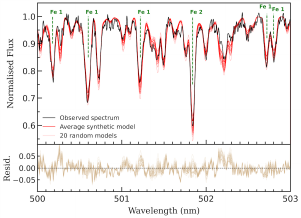}
        
        \caption{An example of the fit between the optimum \ispec{} spectrum and a TIC 48227288A spectrum observed by \minaus{}-Australis. The bold red line is the best fit model spectra. 20 model spectra drawn from random sampling of the model parameter distributions are shown as fainter red lines.}  
        
        \label{fig: TIC48227288 spectrum}
        
\end{figure}

\begin{figure}
\centering
\includegraphics[width=1\columnwidth]{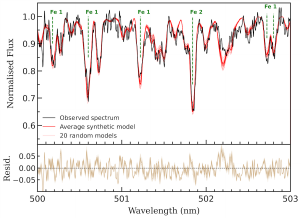}
        
        \caption{An example of the fit between the optimum \ispec{} spectrum and a TIC 339607421A spectrum observed by \minaus{}-Australis. The bold red line is the best fit model spectra. 20 model spectra drawn from random sampling of the model parameter distributions are shown as fainter red lines.}  
        
        \label{fig: 339607421 spectrum}
        
\end{figure}

The initial parameter estimates were then used to generate a synthetic spectrum which enabled a better normalization of the original \minaus{} spectra to be performed before \Teffprim{}, \loggprim{}, \MHprim{}, \vsiniprim{} and \fehprim{} were re-estimated. This procedure was repeated across the 10 selected spectra.  Final stellar parameters were taken as the mean of the individual estimates obtained for each spectra, with standard deviations adopted as uncertainties.  The resulting model spectra provided excellent fits to the observed \minaus{} data, (see Figures~\ref{fig: TIC48227288 spectrum} and \ref{fig: 339607421 spectrum}).

To further refine the primary star characteristics, the averaged \ispec{} outcomes were used as priors in the spectral energy distribution (SED) fitting program, \ariadne{}\footnote{\href{https://github.com/jvines/astroARIADNE}{https://github.com/jvines/astroARIADNE}} \citep{Vines_2022:1420}, along with some additional photometric and astrometric data.  The additional input included Gaia DR3 parallax and $G$, $G_R$, and $G_{RP}$ magnitudes \citep{gaia_2023:1561}; TYCHO $B_{T}$ and $V_{T}$ magnitudes \citep{Hog_2000:291}; 2MASS $J$, $H$, $K$ magnitudes \citep{Cutri_2003:1565}; WISE $W_1$ and $W_2$ magnitudes \citep{Cutri_2021:1566}; and APASS Johnson B and V magnitudes \citep{Henden_2015:1567}. Magnitude data flagged as potentially contaminated, (e.g., by the diffraction spikes from a neighboring stars), were excluded from the fit.

Normal priors were adopted for  \Teffprim{}, \MHprim{} and \loggprim{} based on the \ispec{} results, while default priors were employed for distance \citep[as estimated from Gaia EDR3,][]{Bailer-Jones_2021:1554}, radius (uniform prior from 0.5 to 20 \Rsun) and line of sight extinction, Av, (uniform prior from 0 to the maximum line-of-sight value). Stellar properties were derived from Bayesian model averaging across four atmospheric models:  BT-Settl \citep{Schlafly_2011:1555} , Phoenix V2 \citep{Husser_2013:1556}, Kurucz \citep{Kurucz_1993:1557} and Castelli/Kurucz \citep{castelli_2003:1559}. For TIC 339607421A however, the Kurucz model yielded significantly discrepant results and was therefore excluded from the average.

\begin{figure}
\centering
\includegraphics[width=1\columnwidth]{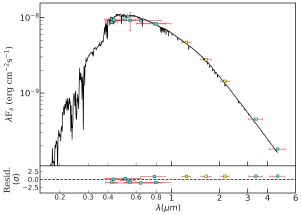}
          
          \caption{Top: TIC 48227288A best fit SED obtained from the \ariadne{} analysis. Black line is the best model fit. Blue squares show photometry data used in determining the SED profile while yellow squares show data excluded from the fitting process due to possible quality issues. Horizontal bars correspond to the filter band-pass width while vertical error bars reflect magnitude uncertainties. Bottom: Residuals relative to flux uncertainties.}
          
          \label{fig: TIC48227288 SED}
          
\end{figure}

\begin{figure}
\centering
\includegraphics[width=1\columnwidth]{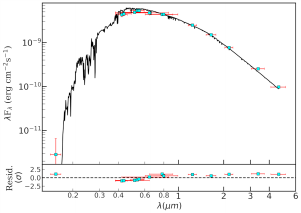}
       
       \caption{Top: TIC 339607421A best fit SED obtained from the \ariadne{} analysis. Black line is the best model fit. Blue squares show photometry data used in determining the SED profile. Horizontal bars correspond to the filter band-pass width while vertical error bars reflect magnitude uncertainties. Bottom: Residuals relative to flux uncertainties.}
       
       \label{fig: TIC 339607421 SED}
       
\end{figure}

Final primary star parameters are reported in Table \ref{table:Host properties}. The model SED fits based on the Castelli/Kurucz atmospheric model are shown in Figures ~\ref{fig: TIC48227288 SED} and \ref{fig: TIC 339607421 SED} for TIC 48227288A and TIC 339607421A, respectively. Although the 2MASS J,H,K data for TIC 48227288A  were excluded from the analysis due to potential contamination, comparison with the model SED showed good agreement (see yellow points in Fig. \ref{fig: TIC48227288 SED}).

Our analysis suggests that TIC 48227288A is a main sequence F3 class star with a radius of $1.61\pm0.03$~\Rsun{}, mass of $1.36{^{+0.06}_{-0.08}}$~\Msun{}, effective temperature of $6723{^{+104}_{-97}}$~K, and close to solar metallicity, \fehprim{}=${-0.03^{+0.06}_{-0.07}}$, with an estimated age of $\sim$2-3 Gyr. TIC 339607421A, on the other hand, is a slightly metal poor, \fehprim{}=$-0.15\pm{0.03}$, main sequence F6 star with a radius of $1.21^{+0.03}_{-0.02}$~\Rsun, mass of $1.09\pm{0.04}$ \Msun{} with an effective temperature of $6402{^{+39}_{-65}}$~K, and likely aged  less than 5 Gyr. These results are consistent, within uncertainties, with the TIC 48227288 and TIC 339607421 parameters listed in version 8.2 of the \textit{TESS} input catalog  \citep{Stassun_2019:965} and as such, were the parameters we used when characterizing the companion stars in each system.


%


\begin{table*}
\renewcommand{\arraystretch}{1.4}
\small
\captionsetup{justification=centering}  
\caption{Stellar properties for TIC 48227288A and TIC 339607421A used in this study. Notes: $\vdag{}$ preferred solution; $\vddag$ excluded from SED fitting process.} 

\label{table:Host properties}      
\centering                          
\begin{tabular}{l c c c }        

\hline\hline                 

Property & TIC 48227288A & TIC 339607421A & Sources \\  

\Xhline{3\arrayrulewidth} 
\noalign{\vskip 3px}


\textbf{Astrometric properties} &                                 &                                & \\
R.A. (hh:mm:ss)                 & 03:04:50.27                     & 03:04:50.27                    & Gaia DR3\\
Decl. (dd:mm:ss)                & 14:51:40.29                     & -52:33:30.85                   & Gaia DR3 \\
$\mu_{\alpha}$ (\masyr)         & $-58.409 \pm 0.028$             & $12.560 \pm 0.013$             & Gaia DR3 \\
$\mu_{\delta}$ (\masyr)         & $-72.082 \pm 0.024$             & $52.692 \pm 0.017$             & Gaia DR3 \\
Parallax (mas)                  & $9.9775 \pm 0.0249$             & $10.0183 \pm 0.0132$           & Gaia DR3 \\
Distance (pc)                   & $100.0473^{+0.2570}_{-0.2584}$$\vdag{}$  & $99.6676^{+0.1221}_{-0.1264}$$\vdag{}$  & Gaia DR3 \\
                                & $100.0712^{+0.1839}_{-0.1401}$  & $99.8724^{+0.1255}_{-0.0932}$  & \ariadne{}, Sect. \ref{subsection:host star analyses} \\
                           
\textbf{Photometric properties} &                                 &                                &          \\
B$\mathrm{_{T}}$ (mag)          & $8.761 \pm 0.017$               & $9.524 \pm 0.020$              & Tycho    \\
V$\mathrm{_{T}}$ (mag)          & $8.273 \pm 0.012$               & $8.998 \pm 0.016$              & Tycho    \\
J (mag)                         & $7.292 \pm 0.020 \vddag$        & $7.978 \pm 0.024$              & 2MASS    \\
H (mag)                         & $7.083 \pm 0.024 \vddag$        & $7.742 \pm 0.069$              & 2MASS    \\
K (mag)                         & $7.030 \pm 0.029 \vddag$        & $7.705 \pm 0.018$              & 2MASS    \\
TESS (mag)                      & $7.778 \pm 0.006$               & $8.466 \pm 0.006$              & TIC v8.2    \\
WISE$\mathrm{_{1}}$ (mag)       & $6.974 \pm 0.035$               & $7.604 \pm 0.029$              & WISE    \\
WISE$\mathrm{_{2}}$ (mag)       & $6.981 \pm 0.019$               & $7.660 \pm 0.020$              & WISE    \\
Gaia (mag)                      & $8.117 \pm 0.003$               & $8.824 \pm 0.003$              & Gaia DR3    \\
Gaia$\mathrm{_{BP}}$ (mag)      & $8.346 \pm 0.003$               & $9.069 \pm 0.003$              & Gaia DR3    \\
Gaia$\mathrm{_{RP}}$ (mag)      & $7.721 \pm 0.004$               & $8.415 \pm 0.004$              & Gaia DR3    \\
Johnson V (mag)                 & $8.211 \pm 0.014$               & $8.921 \pm 0.017$              & ACSS v3    \\
Johnson B (mag)                 & $8.699 \pm 0.018$               & $9.449 \pm 0.018$              & ACSS v3   \\

\textbf{Spectroscopic properties}   &                                 &                                &          \\
Spectral type                       & F5                              &  F8                            & HD catalogue \\
                                    & F3V$\vdag{}$                    &  F6V$\vdag{}$                  & \ariadne{}, Sect. \ref{subsection:host star analyses} \\
\Teffprim{} (K)                     & $6816 \pm 134$                  & $6327 \pm 128$                 & TIC v8.2    \\
                                    & $6723^{+104}_{-97}$$\vdag{}$    & $6402^{+39}_{-65}$$\vdag{}$    & \ariadne{}, Sect. \ref{subsection:host star analyses} \\
\loggprim{} (cgs)                   & $4.17 \pm 0.09$                 & $4.34 \pm 0.08$                & TIC v8.2    \\
                                    & $4.16 \pm 0.02$                 & $4.31 ^{+0.02}_{-0.01}$        & \ariadne{}, Sect. \ref{subsection:host star analyses} \\
Metallicity, \MHprim{}              & $-0.09 \pm 0.11$                & $-0.27 \pm 0.05$               & \ispec{}, Sect. \ref{subsection:host star analyses} \\
Metallicity, \fehprim{}             & $-0.03^{+0.06}_{-0.07}$         & $-0.15 \pm 0.03$               & \ariadne{}, Sect. \ref{subsection:host star analyses} \\
\vsiniprim{} (\kms)                 & $20.0 \pm 1.7$                  & $24.6 \pm 0.6$                 & \ispec{}, Sect. \ref{subsection:host star analyses} \\
                                    & $19.0 \pm 0.9 $                 & $26.0\pm 2.2$                  & \allesfitter{}, Sect. \ref{subsection:Allesfitter analysis} \\
                                    & $17.9 \pm 0.2 $$\vdag{}$        & $24.9\pm 0.2$$\vdag{}$         & Reloaded RM, Sect. \ref{subsection:Reloaded RM method} \\                           
\textbf{Derived stellar properties} &                             &                                &          \\
$M_A$ (\Msun)                       & $1.47 \pm 0.11$             & $1.25 \pm 0.18$                & TIC v8.2    \\
                                    & $1.36^{+0.06}_{-0.08}$$\vdag{}$      & $1.09 \pm 0.04$$\vdag{}$         & \ariadne{}, Sect. \ref{subsection:host star analyses} \\
$R_A$ (\Rsun)                       & $1.64 \pm 0.07$             & $1.25 \pm 0.05$                & TIC v8.2    \\
                                    & $1.61 \pm 0.03\vdag{}$             & $1.21 ^{+0.03}_{-0.02}$$\vdag{}$                & \ariadne{}, Sect. \ref{subsection:host star analyses} \\
$\rho_A$ (cgs)                      & $0.47 \pm 0.11$           & $0.89 \pm 0.20$              & TIC v8.2    \\
                                    & $0.46 \pm 0.02$$\vdag{}$   & $0.88^{+0.04}_{-0.03}$$\vdag{}$      & \ariadne{}, Sect. \ref{subsection:host star analyses} \\
Age (Gyr)                           & $2.3^{+0.8}_{-0.6}$      & $4.3^{+1.0}_{-4.2}$         & \ariadne{}, Sect. \ref{subsection:host star analyses} \\


\noalign{\vskip 6px} 
\hline
\noalign{\vskip 8px} 
\end{tabular}
\\
\justifying
Source references: 2MASS \citep{Cutri_2003:1565};  \allesfitter{} \citep{Gunther_2021:1318};  ACSS v3 \citep{Kharchenko_2001:1705}; \ariadne{} \citep{Vines_2022:1420}; GAIA DR3 \citep{gaia_2023:1561}; HD catalogue \citep{Cannon_1993:991}; \ispec{} \citep{blanco_2014:1396}; TIC~v8.2 \citep{Stassun_2019:965}; Tycho \citep{Hog_2000:291};   WISE \citep{Cutri_2021:1566};  \\
\justifying
\end{table*} 



\subsection{Allesfitter joint analysis of light curve and RV data } \label{subsection:Allesfitter analysis}
To determine the properties of the companion stars and the stellar obliquities of the host stars, we first performed a joint analysis of the \textit{TESS} photometric and \minaus{} spectroscopic data using the \allesfitter{} package \footnote{\href{https://github.com/MNGuenther/allesfitter}{https://github.com/MNGuenther/allesfitter}} \citep{Gunther_2019:1563,Gunther_2021:1318}. This analysis incorporated the host star properties determined earlier from the \ispec{} and \ariadne{} modeling. 

 The priors used in the \allesfitter{} joint analyses are listed in Tables \ref{table:TIC 48227288 Allesfitter_results}, \ref{table:TIC 339607421 Allesfitter_results}, \ref{table:additional TIC 48227288 fit results} and \ref{table:TIC 339607421 additional fit results}. Generous uniform priors were applied to all parameters except for \vsini{} where a 5$\sigma$ (weak) normal prior was employed that was centered on \vsini{} estimates obtained from the \ispec{} analysis of the \minaus{} spe, (\norm{}(20.0,8.5)~\kms{} and \norm{}(24.6,3.0) \kms{} for TIC~48227288  and TIC 339607421 respectively). Boundary conditions for several uniform priors, (mid eclipse time ($T_{0,B}$), ratio of the sum of star radii to semi major axis$(R_A + R_B) / a_B$), primary to secondary stellar radius ratio ($R_B / R_A$), cosine of the angle of orbital inclination ($\cos{i_B}$), orbital period ($P_B$), eccentricity ($\sqrt{e_B} \cos{\omega_B}$, $\sqrt{e_B} \sin{\omega_B}$) and radial velocity semi-amplitude ($K_B$)), were guided by exploratory modeling of the radial velocity and light curve data performed using \radvel\footnote{\href{https://github.com/California-Planet-Search/radvel}{https://github.com/California-Planet-Search/radvel}} \citep{Fulton_2018:274} and the online version of \exofast\footnote{\href{https://astroutils.astronomy.osu.edu/exofast/exofast.shtml}{https://astroutils.astronomy.osu.edu/exofast/exofast.shtml}} \citep{Eastman_2013:299}. Uniform prior ranges for light curve flux uncertainties ($\ln{\sigma_{F}}$) and radial velocity jitter ($\ln{\sigma_\mathrm{jitter}}$) were informed by reported uncertainties in the \textit{TESS} flux and \minaus{} datasets. Following the recommendations of \citet{Espinoza_2015:1704}, the priors for the transformed limb darkening coefficients ($q_1, q_2$) were set to \uni(0,1) as was the specific surface brightness ratios ($J_B$). A wide ranging prior was set for the sky projected obliquity, $\lambda_A$, \uni(-180,180).

It is evident in Figures \ref{fig:TIC48227288 LC} and \ref{fig:TIC339607421 LC} that there is considerable baseline flux variability between eclipses. We modeled the variability as a combination of three effects outlined in \citet{Shporer_2017:1411}: (a)  ellipsoidal distortion of the primary star induced by the tidal bulge raised on the primary surface by the close orbiting companion;  (b) Doppler boosting (or beaming) modulations in which the observed flux is modulated by, for example, radial velocity induced Doppler shifting of the host spectrum; and (c) atmospheric brightness modulation, arising from reflected light and thermal emission from the companion. \allesfitter{} incorporates these effects using the parameters $A_{\mathrm{ellispoidal}}$, $A_{\mathrm{beaming}}$, and  $A_{\mathrm{atmospheric}}$, which define the amplitudes of each component. To test the impact of the baseline variability on modeling outcomes, we carried out additional modeling using light curve data from which the between-eclipse modulation had been removed  with a Savitzky–Golay filter, (referred to hereafter  as ’flat-LC’ data). 

Initial modeling of the TIC 48227288 system revealed significant structure in the RV residuals. To address this, we modeled the RV baseline red noise using a Gaussian process (GP) with a real kernel and broad uniform priors on both the amplitude and timescale parameters (see Table \ref{table:additional TIC 48227288 fit results}). For TIC 339607421, the RV baseline was modeled using a simple offset, again with wide uniform priors (Table \ref{table:TIC 339607421 additional fit results}).

The joint posterior distributions were sampled using the \allesfitter{} dynamic, random walk Nested Sampling (NS) option, employing the \dynesty{} package \citep{Speagle_2020:1608} with a set of 500 live points in the parameter space to explore the posterior distribution. The NS fits were run until a convergence threshold of $\Delta$lnZ$\leq$0.01 was reached, (where Z represents the Bayesian evidence determined during each iteration).



\begin{table*}
\renewcommand{\arraystretch}{1.4}
\small
\captionsetup{justification=centering}  
\caption{Median and 68\% confidence intervals of astrophysical parameters derived for TIC 48227288 by \allesfitter{}. Priors are shown as uniform \uni(a,b) or normal \norm$(\mu,\sigma)$. Additional parameters derived for the eclipse and radial velocity fits can be found in Table \ref{table:additional TIC 48227288 fit results}.Notes:$\vdag{}$ preferred solution. $\vddag{}$ prior not used for flattened light curve model}

\label{table:TIC 48227288 Allesfitter_results}   
\centering                          
\begin{tabular}{l l c c}        

\hline\hline                 

Parameter & Prior & Best Fit$\vdag{}$ & Best Fit (flattened light curve) \\  
\Xhline{3\arrayrulewidth} 

\multicolumn{4}{l}{\textbf{Fitted Parameters}} \\


Radius ratio, $R_{B} / R_{A}$ &                         \uni$(0.3,0.4)$ &      $0.3756\pm0.0011$ &                   $0.3747\pm0.0009$\\
Radii sum to semi-major axis, $(R_{A} + R_{B}) / a$ &   \uni$(0.1,0.3)$ &      $0.2096_{-0.0010}^{+0.0007}$ &        $0.2101\pm0.0005$\\
Cosine of inclination , $\cos{i_B}$ &   \uni$(0,0.4)$ &        $0.1032_{-0.0010}^{+0.0009}$ &        $0.1032_{-0.0007}^{+0.0006}$\\
Mid transit time $T_{0;B}$ (\BJDTDB-2459000) &          \uni$(424.0,426.0)$ &  $395.99372\pm0.00002$ &               $395.99380\pm0.00001$\\
Orbital period , $P_B$  (d) &                           \uni$(2.796,2.996)$ &  $2.8963969\pm0.0000001$ &             $2.8963965\pm0.0000001$\\
$\sqrt{e_B} \cos{\omega_B}$ &                           \uni$(-0.5,0.5)$ &     $-0.027_{-0.003}^{+0.007}$ &          $-0.019_{-0.005}^{+0.004}$\\
$\sqrt{e_B} \sin{\omega_B}$ &                           \uni$(-0.5,0.5)$ &     $0.021_{-0.029}^{+0.022}$ &           $0.047_{-0.016}^{+0.013}$\\
Radial velocity semi-amplitude, $K_B$ (\kms) &          \uni$(53.0,63.0)$ &    $58.62_{-0.14}^{+0.13}$ &             $58.47_{-0.15}^{+0.14}$\\
Surface brightness ratio, $J_B$ &                       \uni$(0,1)$ &          $0.1375_{-0.0007}^{+0.0006}$ &        $0.1417\pm0.0004$\\
Doppler beaming amplitude, $A_\mathrm{beaming}$ (ppt) & \uni$(0,1) \vddag{}$ &          $0.221_{-0.012}^{+0.011}$ &           .....................\\
Atmospheric amplitude, $A_\mathrm{atmosphere}$ (ppt) &  \uni$(0,1) \vddag{}$ &          $0.011_{-0.008}^{+0.012}$ &           .....................\\
Ellipsoidal amplitude, $A_\mathrm{ellipsoidal}$ (ppt) & \uni$(0,5) \vddag{}$ &          $2.915\pm0.026$ &                     .....................\\
Stellar rotational velocity, \vsiniprim{} (\kms) &      \norm$(20,8.5)$ &      $19.0\pm0.9$ &                        $19.2\pm0.9$\\
Sky projected obliquity, $\lambda$ (deg) &              \uni$(-180,180)$ &     $-17.8_{-2.0}^{+1.9}$ &               $-17.5_{-2.2}^{+2.1}$\\
\hline
\multicolumn{4}{l}{\textbf{Derived Parameters}} \\

Primary radius to semi-major axis, $R_A/a$ &            ..................... & $0.1524_{-0.0007}^{+0.0005}$ &        $0.1528\pm0.0003$\\
Mass ratio, $q$ &                                       ..................... & $0.469\pm0.013$ &                     $0.469\pm0.013$\\
Companion radius, $R_\mathrm{B}$ (\Rsun) &              ..................... & $0.605\pm0.011$ &                     $0.604\pm0.011$\\
Companion mass, $M_\mathrm{B}$ (\Msun) &                ..................... & $0.635\pm0.037$ &                     $0.635_{-0.038}^{+0.037}$\\
Companion density, $\rho_\mathrm{B}$ (\cgs) &           ..................... & $4.04_{-0.38}^{+0.42}$ &              $4.07_{-0.38}^{+0.42}$\\
Eccentricity, $e_\mathrm{B}$ &                          ..................... & $0.0013_{-0.0003}^{+0.0011}$ &        $0.0026_{-0.0010}^{+0.0013}$\\
Argument of periastron, $w_\mathrm{B}$ (deg) &          ..................... & $142.5_{-27.7}^{+52.9}$ &             $111.8_{-7.6}^{+15.5}$\\
Orbital semi-major axis, $a$ (AU) &                     ..................... & $0.04916_{-0.00093}^{+0.00094}$ &     $0.04901\pm0.00092$\\
Orbital inclination, $i_\mathrm{B}$ (deg) &             ..................... & $84.08_{-0.05}^{+0.06}$ &             $84.07\pm0.04$\\
Primary transit depth (ppt) $\delta_{\mathrm{tra:TESS}}$ & ..................... & $134.41_{-0.10}^{+0.08}$ &            $134.54\pm0.07$\\
Primary transit impact parameter, $b_\mathrm{tra;B}$ &  ..................... & $0.676\pm0.004$ &                     $0.674\pm0.003$\\
Primary transit total duration, $T_\mathrm{tot;B}$ (d) & ..................... & $0.1700_{-0.0006}^{+0.0004}$ &        $0.1702\pm0.0003$\\
Primary transit full duration, $T_\mathrm{full;B}$ (d) & ..................... & ..................... &               .....................\\
Secondary transit depth (ppt) $\delta_{\mathrm{occ:TESS}}$ & ..................... & $18.55_{-0.07}^{+0.08}$ &             $18.97\pm0.05$\\
Secondary transit epoch, $T_\mathrm{0;occ;B}$ (\BJDTDB-2459000) & ..................... & $394.54374\pm0.00015$ &               $394.54384\pm0.00011$\\
Secondary transit impact parameter, $b_\mathrm{occ;B}$ & ..................... & $0.678_{-0.005}^{+0.004}$ &           $0.677_{-0.004}^{+0.003}$\\


\noalign{\vskip 6px} 
\hline
\noalign{\vskip 8px} 
\end{tabular}
\\
\justifying

\end{table*} 



\subsubsection{TIC 48227288: Allesfitter Joint Analysis Results}

\begin{figure}
\resizebox{\hsize}{!} 
{\includegraphics[]{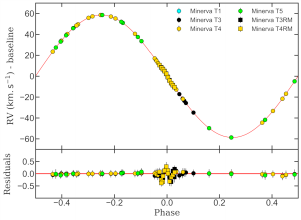}} 
        
         \caption{Phase folded radial velocity model for TIC 48227288 obtained from the \allesfitter{} analysis. Observations from each of the \minaus{} telescopes are shown including observations made during the primary eclipse. The red line shown represents the best fit model. Residual velocities to the best fit model are shown in the bottom panel.}
  
         \label{fig: TIC48227288 RV fit}
     
\end{figure}

\begin{figure*}
\resizebox{\hsize}{!}
{\includegraphics[]{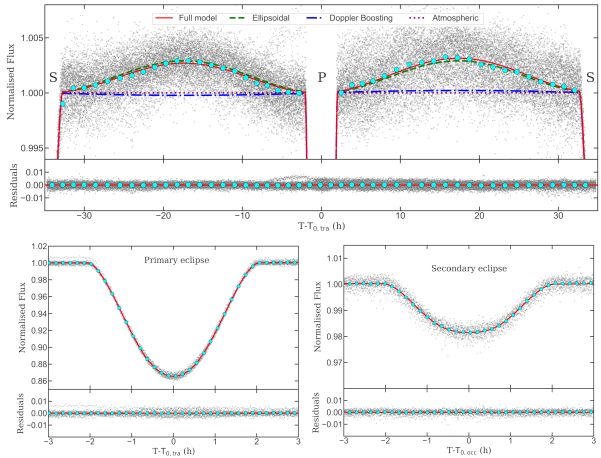}}
        
         \caption{Phase folded TIC 48227288 \textit{TESS} PDCSAP data showing the model fit to the flux baseline modulation (top), primary eclipse (bottom left) and secondary eclipse (bottom right). The positions of the primary and secondary eclipses in the top panel are marked by 'P' and 'S'. \textit{TESS} observations are shown as gray points. Cyan circles are \textit{TESS} observations binned with a cadence of 10 minutes (for both eclipses) and 90 minutes (out of eclipse data). Included in the phase curve model fit shown in the top graph are the ellipsoidal (green dashed line), atmospheric (purple dotted line) and beaming modulations( blue dash-dotted line). The combined median model fit is plotted as a solid red line in all plots. Residuals to the combined model are show in the bottom panels of each plot.}
         
         \label{fig: TIC48227288 LC fit}
         
\end{figure*}

The results of the joint analysis performed for the TIC 48227288 system are listed in Tables \ref{table:TIC 48227288 Allesfitter_results} and \ref{table:additional TIC 48227288 fit results}. Comparison of the original model results using the unfiltered light curve data with flat-LC model outcomes reveals only minor differences that, for the most part, fall within the stated parameter uncertainties. More significant discrepancies among the principal parameters are limited to the orbital period, surface brightness and secondary eclipse depth where differences are nonetheless small, ($\Delta$P= 0.03 s, $\Delta J_B$ = 0.004 and $\Delta\delta_{\mathrm{occ:TESS}}$ = 0.4~ppt). This suggests that the baseline flux modulation was well captured in the original modeling. We therefore adopt the original model findings as our preferred \allesfitter{} modeling solution for the TIC 48227288 system. Fits of this model to the RV and photometric data are shown in Figures \ref{fig: TIC48227288 RV fit} and \ref{fig: TIC48227288 LC fit}.

The companion TIC 48227288B was found to have a radius of $0.605\pm0.011$ \Rsun{} and mass $0.635\pm0.037$~\Msun{} consistent with a late K-class star. The companion follows a short period, essentially circular orbit (a=$0.0492\pm0.0009$ AU, P=2.8963969$\pm0.0000001$ days, e=$0.0013_{-0.0003}^{+0.0011}$). The orbital inclination is well constrained at $84.08^{+0.06}_{-0.05}$$^\circ$. Both primary and secondary eclipses are well fit by the global model (Figure \ref{fig: TIC48227288 LC fit}, bottom panels) with depths of $134.4\pm 0.1$ and $18.6 \pm 0.1$ ppt, respectively. The system geometry is such that the eclipses are nearly total, with approximately 96\% of the companion disk occluded during the secondary eclipse. While the baseline flux shows considerable modulation between eclipses, (Figure \ref{fig: TIC48227288 LC fit}, top panel), the inclusion of  ellipsoidal variation, Doppler beaming, and atmospheric modulation terms appears to have adequately accounted for this.

\begin{figure}
\resizebox{\hsize}{!}
{\includegraphics[]{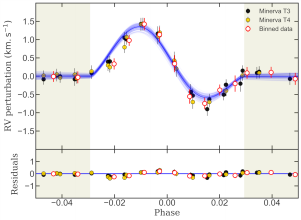}}
        
         \caption{Phase folded radial velocity model of the Rossiter-McLaughlin effect for TIC 48227288 formed after subtraction of the orbital radial velocities. Observations from the two \minaus{} telescopes used during eclipse events are shown as well as binned data to guide the eye. The white area marks the transit window. Blue lines show 100 randomly drawn posterior radial velocity models. Residual velocities to the best fit model are shown in the bottom panel.}
         
         \label{fig: TIC48227288 RM fit}
         
\end{figure}

The principal aim of this study is to constrain the stellar obliquity of the primary star. The large RV semi-amplitude exhibited by the TIC 48227288 system, (K$_B$=$58.62^{+0.13}_{-0.14}$ \kms), renders the RM effect difficult to discern in Figure \ref{fig: TIC48227288 RV fit}.  However, the RM perturbation is clearly evident after removing the orbital RV component and isolating the eclipse event, (Figure \ref{fig: TIC48227288 RM fit}). Its slight asymmetry suggests a low but non-zero obliquity, which is supported by the findings of the global model which indicates that TIC 48227288B follows a marginally misaligned, prograde orbit with a sky projected obliquity of $-17.8^{+1.9}_{-2.0}$$ ^{\circ}$.


\subsubsection{TIC 339607421: Allesfitter Joint Analysis Results}

\begin{figure*}
\resizebox{\hsize}{!}
{\includegraphics[]{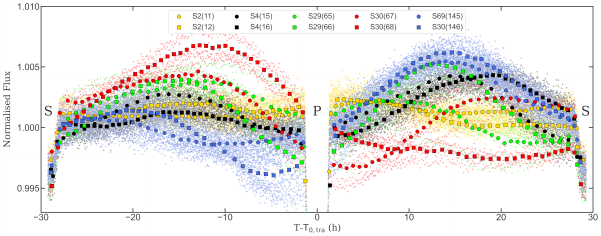}}
        
         \caption{Phase folded TIC 339607421 \textit{TESS} PDCSAP data showing the variability in the baseline behavior between eclipses. The positions of the primary and secondary eclipses in the top panel are marked by 'P' and 'S'. \textit{TESS} observations are shown as smaller points color coded with respect to the observation sector. Larger points show data binned to a cadence of $\sim$60 minutes. Binned data have been grouped not only by sector number but also orbit number (shown in brackets in the legend). The complex behavior of the baseline over multiple timescales is clearly evident.}
         
         \label{fig:TIC339607421 LC fit}
         
\end{figure*}

\begin{figure*}
\resizebox{\hsize}{!}
{\includegraphics[]{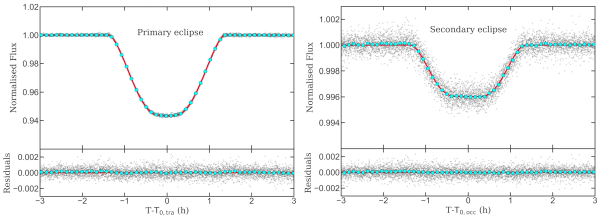}}
        
         \caption{\allesfitter{} model fit to phase folded TIC 339607421 \textit{TESS} PDCSAP data that has been filtered to remove baseline modulation. Left: Model fit to the primary eclipse. Right: Model fit to the secondary eclipse.  \textit{TESS} observations are shown as gray points. Cyan circles are \textit{TESS} observations binned with a cadence of $\sim$6 minutes. The best fit model fit is plotted as a solid red line. Residuals are show in the bottom panels of each plot}
         
         \label{fig:TIC339607421 LC flat fit}
         
\end{figure*}

\begin{figure}
\resizebox{\hsize}{!}
{\includegraphics[]{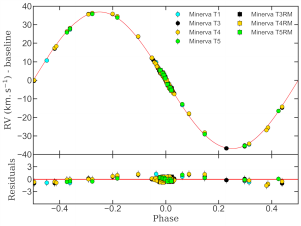}}
        
         \caption{Phase folded radial velocity model for TIC 339607421 obtained from the \allesfitter{} analysis. Observations from each of the \minaus{} telescopes are shown including observations made during the primary eclipse. The red line shown represents the best fit model. Residual velocities to the best fit model are shown in the bottom panel.}
         
         \label{fig:TIC339607421 RV flat fit}
         
\end{figure}

The results of the \allesfitter{} joint analysis performed for the TIC 339607421 system are listed in Table \ref{table:TIC 339607421 Allesfitter_results} with additional results provided in Table \ref{table:TIC 339607421 additional fit results}. As is evident in Figure \ref{fig:TIC339607421 LC fit}, the variation in the TIC 339607421 light curve baseline flux  is notably more complex than that observed for TIC 48227288, with modulations occurring over a greater number of timescales. While efforts to fully model this baseline variation within the Allesfitter framework achieved only limited success, the modeling outcomes from both the original and flat-LC datasets are consistent within the stated uncertainties (see Table \ref{table:TIC 339607421 Allesfitter_results}). Nevertheless, given the complications involved in modeling the between-eclipse baseline we adopt the flat-LC model as the preferred \allesfitter{} modeling solution for the TIC 339607421 system. Fits of the flat-LC model to the photometric and RV data are shown in Figures \ref{fig:TIC339607421 LC flat fit} and \ref{fig:TIC339607421 RV flat fit} respectively. 

The companion TIC 339607421B has a radius R$_B=0.291\pm0.006$~\Rsun{} and mass M$_B=0.294\pm 0.013$~\Msun{}) consistent with an M3-M4 class star. It follows an essentially circular orbit, (e=$0.0010_{-0.0004}^{+0.0011}$), with a period of $2.43821243\pm 0.00000002$ days, a semi-major axis of $0.0391\pm0.0008$ AU, and orbital inclination of $83.93\pm0.03$$^{\circ}$ . As expected for a smaller, cooler companion, the eclipse depths are more modest than in the TIC 48227288 system with primary and secondary eclipse depths of $56.66\pm0.02$ ppt and $3.98\pm0.02$ ppt respectively. In contrast to TIC 48227288, the TIC 339607421 system experiences a full secondary eclipse, albeit briefly, with the companion completely occluded for $\sim$32 minutes during the secondary eclipse.

\begin{figure}
\resizebox{\hsize}{!}
{\includegraphics[]{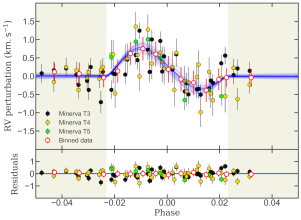}}
        
         \caption{Phase folded radial velocity model of the Rossiter-McLaughlin effect for TIC 339607421 formed after subtraction of the orbital radial velocities. Observations from the three \minaus{} telescopes used during eclipse events are shown as well as binned data to guide the eye. The white area marks the transit window. Blue lines show 100 randomly drawn posterior radial velocity models. Residual velocities to the best fit model are shown in the bottom panel.}
         
         \label{fig:TIC339607421 RM flat fit}
         
\end{figure}

The RM effect was observed over three eclipses for this system. As can be seen in Figure \ref{fig:TIC339607421 RM flat fit}, the RM signature is less pronounced than in the TIC 48227288 system due to the smaller size of the companion, (the RM amplitude is approximately 600\ms{} compared to 830\ms{} for TIC 48227288). The RM perturbation is nearly symmetric, indicating a low stellar obliquity which is confirmed by the global analysis which yields a sky projected obliquity, $\lambda = -14.7_{-5.9}^{+5.4}\degree$. The combination of low median obliquity and larger associated uncertainties suggests that any misalignment in the TIC 339607421 system is likely to be minor.



\begin{table*}
\renewcommand{\arraystretch}{1.4}
\small
\captionsetup{justification=centering}  
\caption{Median and 68\% confidence intervals of astrophysical parameters derived for TIC 339607421 by \allesfitter{}. Priors are shown as uniform \uni(a,b) or normal \norm$(\mu,\sigma)$. Additional parameters derived for the eclipse and radial velocity fits can be found in Table \ref{table:TIC 339607421 additional fit results}.Notes:$\vdag{}$ preferred solution. $\vddag{}$ prior not used for flattened light curve model}

\label{table:TIC 339607421 Allesfitter_results}   
\centering                          
\begin{tabular}{l l c c}        

\hline\hline                 

Parameter & Prior & Best Fit & Best Fit (flattened light curve)$\vdag{}$ \\  
\Xhline{3\arrayrulewidth} 

\multicolumn{4}{l}{\textbf{Fitted Parameters}} \\


Radius ratio, $R_{B} / R_{A}$ &                         \uni$(0.23,0.25)$ &    $0.2388_{-0.0009}^{+0.0010}$ &        $0.2395_{-0.0006}^{+0.0005}$\\
Radii sum to semi-major axis, $(R_{A} + R_{B}) / a$ &   \uni$(0.17,0.19)$ &    $0.1793_{-0.0007}^{+0.0006}$ &        $0.1788\pm0.0004$\\
Cosine of inclination , $\cos{i_B}$ &   \uni$(0,0.2)$ &        $0.1056_{-0.0009}^{+0.0008}$ &        $0.1058\pm0.0005$\\
Mid transit time $T_{0;B}$ (\BJDTDB-2459000) &          \uni$(282.7,283.0)$ & $282.89461\pm0.00003$ &               $282.89468\pm0.00001$\\
Orbital period , $P_B$  (d) &                           \uni$(2.438,2.439)$ &  $2.43821227\pm0.00000007$ &             $2.43821243\pm0.00000002$\\
$\sqrt{e_B} \cos{\omega_B}$ &                           \uni$(-0.5,0.5)$ &     $0.005_{-0.007}^{+0.008}$ &           $0.017\pm0.005$\\
$\sqrt{e_B} \sin{\omega_B}$ &                           \uni$(-0.5,0.5)$ &     $-0.004_{-0.031}^{+0.029}$ &          $-0.015_{-0.027}^{+0.037}$\\
Radial velocity semi-amplitude, $K_B$ (\kms) &          \uni$(36,38)$ &        $36.84\pm0.14$ &                      $36.85_{-0.12}^{+0.13}$\\
Surface brightness ratio, $J_B$ &                       \uni$(0,1)$ &          $0.0649_{-0.0010}^{+0.0009}$ &        $0.0696_{-0.0003}^{+0.0004}$\\
Doppler beaming amplitude, $A_\mathrm{beaming}$ (ppt) & \uni$(0,5) \vddag{}$ &          $1.153\pm0.009$ &                     .....................\\
Atmospheric amplitude, $A_\mathrm{atmosphere}$ (ppt) &  \uni$(0,5) \vddag{}$ &          $0.497\pm0.020$ &                     .....................\\
Ellipsoidal amplitude, $A_\mathrm{ellipsoidal}$ (ppt) & \uni$(0,5) \vddag{}$ &          $2.051\pm0.020$ &                     .....................\\
Stellar rotational velocity, \vsiniprim{} (\kms) &      \norm$(24.6,3.0)$ &    $25.8_{-2.1}^{+2.2}$ &                $26.0\pm2.2$\\
Sky projected obliquity, $\lambda$ (deg) &              \uni$(-180,180)$ &     $-15.3_{-5.5}^{+5.3}$ &               $-14.7_{-5.9}^{+5.4}$\\

\hline
\multicolumn{4}{l}{\textbf{Derived Parameters}} \\

Primary radius to semi-major axis, $R_A/a$ &            ..................... & $0.1447\pm0.0006$ &                   $0.1442\pm0.0004$\\
Mass ratio, $q$ &                                       ..................... & $0.271\pm0.007$ &                     $0.270\pm0.007$\\
Companion radius, $R_\mathrm{B}$ (\Rsun) &              ..................... & $0.290\pm0.006$ &                     $0.291\pm0.006$\\
Companion mass, $M_\mathrm{B}$ (\Msun) &                ..................... & $0.295_{-0.013}^{+0.014}$ &           $0.294\pm0.013$\\
Companion density, $\rho_\mathrm{B}$ (\cgs) &           ..................... & $17.10_{-1.59}^{+1.72}$ &             $16.90_{-1.55}^{+1.67}$\\
Eccentricity, $e_\mathrm{B}$ &                          ..................... & $0.0005_{-0.0004}^{+0.0014}$ &        $0.0010_{-0.0004}^{+0.0011}$\\
Argument of periastron, $w_\mathrm{B}$ (deg) &          ..................... & $256.6_{-188.7}^{+39.8}$ &            $285.1_{-243.1}^{+33.3}$\\
Orbital semi-major axis, $a$ (AU) &                     ..................... & $0.03898_{-0.00079}^{+0.00085}$ &     $0.03911_{-0.00079}^{+0.00083}$\\
Orbital inclination, $i_\mathrm{B}$ (deg) &             ..................... & $83.94\pm0.05$ &                      $83.93\pm0.03$\\
Primary transit depth (ppt) $\delta_{\mathrm{tra:TESS}}$ & ..................... & $56.56_{-0.08}^{+0.07}$ &             $56.66\pm0.02$\\
Primary transit impact parameter, $b_\mathrm{tra;B}$ &  ..................... & $0.730\pm0.004$ &                     $0.734\pm0.001$\\
Primary transit total duration, $T_\mathrm{tot;B}$ (d) & ..................... & $0.1135\pm0.0003$ &                   $0.1129\pm0.0001$\\
Primary transit full duration, $T_\mathrm{full;B}$ (d) & ..................... & ..................... &               .....................\\
Secondary transit depth (ppt) $\delta_{\mathrm{occ:TESS}}$ & ..................... & $4.18\pm0.05$ &                       $3.98\pm0.02$\\
Secondary transit epoch, $T_\mathrm{0;occ;B}$ (\BJDTDB-2459000) & ..................... & $284.11389_{-2.43845}^{+0.00033}$ &   $284.11462\pm0.00011$\\
Secondary transit impact parameter, $b_\mathrm{occ;B}$ & ..................... & $0.729\pm0.004$ &                     $0.733_{-0.003}^{+0.002}$\\


\noalign{\vskip 6px} 
\hline
\noalign{\vskip 8px} 
\end{tabular}
\\
\justifying

\end{table*} 



\subsection{Reloaded Rossiter McLaughlin Analysis}\label{subsection:Reloaded RM method}
The classical RV based approach to measuring stellar obliquities, such as employed above, relies on modeling perturbations in the CCF centroid location caused by distortions that occur in the stellar lines during eclipse events. However, as noted by \citet{Cegla_2016:368} and \citet{Bourrier_2017:421,BOurrier_2022:1106}, this method can introduce biases if the occulted portion of the stellar line profile is not accurately represented. In response to these limitations, the recently developed RRM technique offers an alternative approach that addresses some of the shortcomings of the more traditional analysis. The RRM method involves extracting the radial velocity ('local RV') of the stellar surface region occulted by the transiting companion. Using the framework of \citet{Cegla_2016:368} it is then possible to use the local RVs to reconstruct the trajectory of the companion's 'shadow' as it traverses the stellar disk, and thereby in turn determine the stellar obliquity.

To independently verify the obliquity estimates obtained from our Allesfitter analysis, we applied the RRM technique to a number of selected Minerva eclipse events (see Table \ref{table:Reloaded RM obs}). Each selected event includes a minimum of four high cadence observations taken either immediately prior to ingress, or shortly following egress. 

The RRM calculation then proceeded as follows. Disk-integrated CCFs generated during the computation of eclipse event RVs were first scaled by their median continuum flux level, (defined here the CCF region > 2$\sigma$ from the computed RV, where $\sigma$ is the Gaussian width of the CCF profile), and CCF uncertainties calculated from the standard deviation of the continuum region. \allesfitter{} light-curve and RV model predictions were then used to further normalize the CCF's to account for loss of light during the eclipse and correct for Keplerian motion.

\begin{figure}
\resizebox{\hsize}{!}
{\includegraphics[]{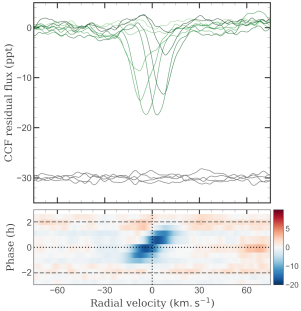}}
        
         \caption{Upper panel: Residual line profiles extracted from \minaus{} observations of the TIC 48227288 eclipse that took place on the night of April 30, 2023. Grey profiles are out-of-transit residuals that have been offset by 30 ppt to improve clarity. Lower panel: Trace plot illustrating the path of the Doppler shadow of the eclipsing body as it traverses the host star disc. Horizontal dashed lines show ingress and egress while the dotted line marks the mid eclipse point. The color bar shows the CCF flux.}
         
         \label{fig: TIC48227288 Reloaded RM residual plot}
         
\end{figure}

A master out-of-transit CCF for each eclipse event was then constructed by averaging all out-of-transit CCFs for that event. A Gaussian profile was fitted to each master CCF and the centroid RV determined and subsequently used to shift all CCFs for that event into the stellar rest frame. Residual line profiles, representing the missing light from the stellar surface blocked by the companion, were then derived by subtracting each in-transit CCF from the corresponding master CCF. As an example, Figure \ref{fig: TIC48227288 Reloaded RM residual plot} presents the residual profiles generated for the TIC 48227288 transit 2a eclipse event. The corresponding trace plot, shown in the bottom panel of the figure, clearly shows the trajectory of the companion 'shadow' as it crosses the stellar disk. Following standard practice in RRM analyses \citep[e.g.,][]{Bourrier_2020:459,Allart_2020:773}, we excluded data points near the stellar limb, where reduced flux and lower signal-to-noise ratios (SNRs) due to partial occultation degraded the reliability of the extracted local RVs.

The remaining in-transit residuals were fitted with Gaussian functions to determine the local RVs. These measurements were then combined across all eclipse events for each system and modeled using the semi-analytic formalism of \citet{Cegla_2016:368}, initially assuming rigid-body stellar rotation. In this model, local RV values depend on the following: the orbital phase ($\phi$), the scaled semi-major axis ($a/R_A$), orbital inclination ($i_{orb}$), radius of the secondary ($R_B$), sky-projected obliquity ($\lambda$), equatorial rotational velocity of the host star (\veqprim{}), and limb darkening (modeled using a quadratic law). Importantly, the stellar inclination, $i_\star$, is fixed at 90° and as such we are unable to determine the true obliquity, $\psi_A$  and the value of \veqprim{} obtained in the model outcomes corresponds to \vsiniprim{}.


 We employed the emcee MCMC sampler \citep{Foreman_2013:574} to estimate the posterior distributions of the model parameters. Gaussian priors for ($a/R_{A}$, $R_{B}$, $i_{orb}$) as well as the quadratic limb darkening coefficients, $U_1$ and $U_2$ were informed by the \allesfitter{} posteriors derived in Section \ref{subsection:Allesfitter analysis} while broad uniform priors were assigned to $\lambda$ and \veqprim{} (see Table \ref{table: Reloaded RM results}). Sampling utilized 50 walkers and continued until the number of steps exceeded $\sim$50 times the autocorrelation length.  Final posterior distributions were computed after discarding an appropriate number of burn-in steps, which were determined visually. 

We also explored an alternative model whereby rotation of the primary was allowed to vary with latitude (differential rotation model). In this framework, we independently sampled the equatorial rotational velocity, \veqprim{}, the stellar inclination $i_\star$, and the relative shear parameter $\alpha$, where $\Omega = \Omega_{eq}(1-\alpha sin^{2}\theta))$, which describes the relative rotation rate between the poles and equator. Again, wide uniform priors were adopted (Table 8). By sampling $i_\star$ independently, we could compute the true three-dimensional obliquity ($\psi$) via:

\begin{equation} \label{equat 1}
\psi=\mathrm{cos}\left(\mathrm{sin}\:i_{\star}\:\mathrm{cos }\:\lambda \: \mathrm{sin} \:i_{p} + \mathrm{cos }\: i_{\star}\: \mathrm{cos }\: i_{p}\right)^{-1}
\end{equation}

\subsubsection{Reloaded Rossiter McLaughlin Results}\label{subsection:Loaded RM results}

\begin{figure*}
\resizebox{\hsize}{!}
{\includegraphics[]{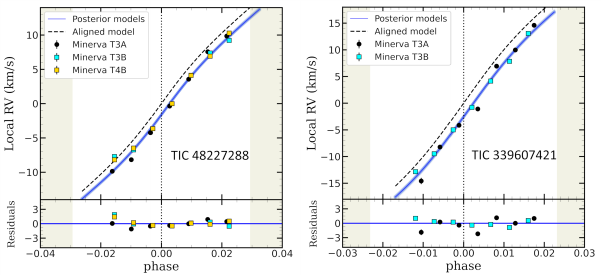}}
        
         \caption{Reloaded RM fit to local radial velocities extracted for TIC 48227288 (left) and TIC 339607421 (right) assuming a rigidly rotating primary. Upper panels show the local RV as a function of phase for multiple eclipse observations accompanied by 50 randomly chosen Reloaded RM model posteriors.  The dashed line marks the position of the best fit $\lambda=0\degree$ model. Lower panels show the local velocity residuals against the best fit model. The transit window is shown as the white area while the vertical dotted line marks the mid transit time.}
         
         \label{fig:Reloaded RM fit to local RV data}
         
\end{figure*}

The results of the RRM analysis for TIC 48227288 and TIC 339607421 are presented in Table \ref{table: Reloaded RM results} and the rigid model fits to the local RVs are shown in Figure \ref{fig:Reloaded RM fit to local RV data}. From the rigid model we derive sky projected obliquities of $\lambda_{A} = -9.5 \pm 0.2\degree$ for TIC 48227288 and $\lambda_{A} = -8.2 \pm 0.2\degree$ for TIC 339607421. These values are lower than those obtained from the joint \allesfitter{} analyses ($-17.8\degree$ and $-14.7\degree$, respectively), but still indicate a minor level of misalignment in both systems, (as evidenced when we compare the position of the $\lambda=0\degree$ model line with the best fit lines in Figure \ref{fig:Reloaded RM fit to local RV data}). The values of \vsiniprim{$_{\star}$} inferred from the rigid RRM model ($17.9  \pm 0.2$~\kms{} for TIC 48227288 and $24.9 \pm 0.2$~\kms{} for TIC 339607421), were consistent within stated uncertainties with values determined from \ispec{} analysis of \minaus{} spectra and \allesfitter{} joint analysis of RV and photometric data (see Table \ref{table:Host properties}). 

When adopting a differential rotation model, (right-hand column of Table \ref{table: Reloaded RM results}), we find sky projected obliquities of $\lambda_{A}$ = -10.5 $\pm0.2\degree$ for TIC 48227288 and $\lambda_{A}$ = -8.8 $\pm0.2\degree$ for TIC 339607421,  broadly consistent with the rigid model results. However, the inferred projected rotational velocities, (\vsiniprim{$_{\star}$}= 12.2 \kms{} and 20.3 \kms{} for TIC 48227288 and TIC 339607421 respectively), are significantly lower than indicated by the \ispec, \allesfitter{} or rigid body RRM modeling. Moreover, the differential rotation parameter, $\alpha$, is poorly constrained and was constantly driven to the lower bounds of our uniform priors even when the lower bound was extended to the unrealistic value of -10. Due to these inconsistencies, we do not consider the differential rotation model results to be physically reliable and therefore disregard them in our subsequent analysis.

We note that the RRM technique has been predominantly applied to exoplanetary systems, and its application to binary star systems remains relatively unexplored \citep[see][]{Kunovac_2020:1209}. As such we were especially cognizant of the warning issued by Kunovac, that even in SB1 binaries such as those considered here, flux from the secondary component can potentially distort the cross-correlation function (CCF), thereby biasing the measurement of local RVs and yielding spurious misalignment signatures. For this to be a significant issue however, the 'secondary' CCF generated during the analysis of the spectra must exceed the noise of the data and this is dependent not only on the light emitted by the secondary but also how the secondary spectra is impacted by convolution with the primary mask that is employed to generate the CCFs. As noted in Section \ref{subsec:Spectroscopy} we used an F3V mask to generate TIC 48227288 CCFs and an F6V mask to generate TIC 339607421 CCFs. To assess the potential impact of the secondary component on our results, we visually inspected the out-of-transit CCFs generated using the aforementioned primary masks — paying particular attention to velocity regions where the secondary signal is expected based on the \allesfitter{} models. In all cases, no secondary CCF was visually apparent suggesting that any contribution from the secondary component was likely suppressed below the general noise level, and thus did not significantly affect our analysis.


\begin{table} 
\small
\renewcommand{\arraystretch}{1.1}
\caption{ Details of \minaus{} observations used in Reloaded RM modeling. }
\label{table:Reloaded RM obs}
\begin{tabular}{lcccc}
\hline\hline

\noalign{\vskip 3px}
 & &  &   \multicolumn{2}{c}{Observations} \\
\cline{4-5}
\noalign{\vskip 3px}
 & Date & Telescope &   Total & Out-of-eclipse \\

\noalign{\vskip 3px}
\hline
\noalign{\vskip 3px}


\textbf{TIC 48227288}  &  &   &   & \\
Transit 1  & Apr 1 2023    & T3 &  16 & 6  \\
Transit 2a & Apr 30 2023   & T3 &  13 & 4  \\
Transit 2b & Apr 30 2023   & T4 &   14 & 4  \\
\textbf{TIC 339607421} &   &   &   \\
Transit 3a & Sept 9 2023   & T3 &  19 & 9  \\
Transit 3b & Sept 9 2023   & T4 &  17 & 7 \\


\noalign{\vskip 3px}
\hline

\end{tabular}

\end{table}


\begin{table*}
\renewcommand{\arraystretch}{1.4}
\small
\captionsetup{justification=centering}  
\caption{Median and 68\% confidence intervals for stellar parameters derived from the RRM analysis of TIC 48227288 and TIC 339607421 eclipse events. Priors are shown as uniform \uni(a,b) or normal \norm$(\mu,\sigma)$. Notes:$\vdag{}$ preferred solution. $\vddag{}$ fixed under rigid model assumption. $^{\ast{}}$ prior not used in rigid model. $^{\ast\ast}$~hard limits at 0,1 also imposed.}

\label{table: Reloaded RM results}   
\centering                          
\begin{tabular}{l l c c}        

\hline\hline                 

Parameter & Prior & Rigid Body & Differential Rotation \\  
\Xhline{3\arrayrulewidth} 

\multicolumn{4}{l}{\textbf{TIC 48227288}} \\


Sky projected obliquity, $\lambda_{A}$ (deg)      & \uni$(-180,180)$          & $-9.5\pm0.2$$\vdag{}$   & $-10.5\pm{+0.2}$ \\
Stellar rotational velocity, \veqprim{} (\kms)    & \uni$(0,100)$             & $17.9\pm0.2$$\vdag{}$     & $13.3\pm 0.2$ \\
Stellar inclination, $i_{\star}$ (deg)            & \uni$(0,180)$$^{\ast{}}$  & 90$\vdag{}$$\vddag{}$     & $66.0_{2.3}^{+62.6}$\\
Differential rotation, $\alpha$                   &\uni$(-1,1)$$^{\ast{}}$  & 0$\vdag{}$$\vddag{}$        & $-0.99_{-0.01}^{+0.02}$\\
Orbital inclination, $i_\mathrm{orb}$ (deg)       & \norm$(84.08,0.04)$       & $84.13\pm0.06$            & $84.14\pm0.06$ \\
Semi-major axis to radius ratio, $a/ R_{A}$       & \norm$(6.563,0.030)$      & $6.569_{-0.029}^{+0.031}$ & $6.572\pm0.030$ \\
Companion Radius, $R_\mathrm{B}$ (\Rsun)          & \norm$(0.605,0.011)$      & $0.599\pm+0.011$          & $0.580_{-0.008}^{+0.006}$ \\
LDC constant , $U_{1}$                            & \norm$(0.16,0.17)^{\ast\ast}$        & $0.06_{-0.04}^{+0.08}$    & $0.06_{-0.05}^{+0.08}$ \\
LDC constant , $U_{2}$                            & \norm$(0.39, 0.26)^{\ast\ast}$       & $0.16_{-0.11}^{+0.17}$    & $0.12_{-0.09}^{+0.14}$ \\
True obliquity, $\psi$ (deg)                      & .....................     &.....................      & $28.9\pm 3.8$ \\
\hline
\multicolumn{4}{l}{\textbf{TIC 339607421}} \\


Sky projected obliquity, $\lambda_A$ (deg)        & \uni$(-180,180)$          & $-8.2\pm0.2$$\vdag{}$     & $-8.8\pm0.2$ \\
Stellar rotational velocity, \veqprim{} (\kms)    & \uni$(0,100)$             & $24.9\pm0.2$$\vdag{}$     & $21.2_{-1.6}^{+1.9}$ \\
Stellar inclination, $i_{\star}$ (deg)            & \uni$(0,180)$$^{\ast{}}$  & 90$\vdag{}$$\vddag{}$     & $72.9_{-6.1}^{+6.5}$ \\
Differential rotation, $\alpha$                   & \uni$(-1,1)$$^{\ast{}}$   & 0$\vdag{}$$\vddag{}$      & $-0.93_{-0.06}^{+0.11}$ \\
Orbital inclination, $i_\mathrm{orb}$ (deg)       & \norm$(83.93,0.03)$       & $83.93\pm0.03$            & $83.93\pm0.03$ \\
Semi-major axis to radius ratio, $a/ R_{A}$       & \norm$(6.932,0.018)$      & $6.932\pm0.018$           & $6.932\pm0.018$\\
Companion Radius, $R_\mathrm{B}$ (\Rsun)          & \norm$(0.291,0.006)$      & $0.291\pm0.006$           & $0.290\pm0.006$  \\
LDC constant , $U_{1}$                            & \norm$(0.30,0.10)^{\ast\ast}$        & $0.35\pm0.10$             & $0.36\pm0.10$  \\
LDC constant , $U_{2}$                            & \norm$(0.02, 0.12)^{\ast\ast}$       & $0.11_{-0.08}^{+0.10}$    & $0.12_{-0.08}^{+0.10}$   \\
True obliquity, $\psi$ (deg)                      & .....................     & .....................     & $14.0_{-4.0}^{+5.1}$ \\

\noalign{\vskip 6px} 
\hline
\noalign{\vskip 8px} 
\end{tabular}
\\
\justifying

\end{table*}


\section{Discussion} \label{Section: Discussion}


\begin{figure}
\resizebox{\hsize}{!} 
        {\includegraphics[]{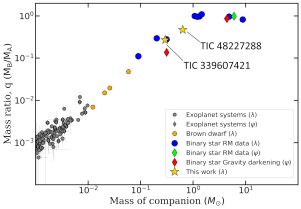}}
  \caption{Log-log plot of the companion mass-mass ratio parameter space for published obliquity studies. In this and remaining obliquity graphs, exoplanetary obliquity studies are shown in gray, while brown dwarfs and binary star studies are colored. Sky projected obliquities, $\lambda$, and true obliquities, $\psi$, are designated by circles and diamonds respectively. The two target systems that are the focus of this study are shown as yellow stars.} 
     \label{fig:obliq vs q,M2}
\end{figure}

We report measurements of the sky-projected stellar obliquities for TIC 48227288 and TIC 339607421, contributing to the relatively limited sample of obliquity measurements available for binary stellar systems (Table \ref{table A1:previous studies}). Sky projected obliquities were derived using both the classical radial velocity (RV) perturbation approach via the \allesfitter{} software package and also using the Reloaded Rossiter–McLaughlin (RRM) technique. The classical method yielded marginally misaligned prograde orbits for the companion stars ($\lambda_{\mathrm{A}} = -17.8^{+1.9}_{-2.0} \degree$ and $\lambda_{\mathrm{A}} = -14.7^{+5.4}_{-5.9} \degree$ for TIC 48227288 and TIC 339607421 respectively). The RRM analysis, assuming a rigidly rotating primary, returned smaller values ($\lambda_{\mathrm{A}} = -9.5 \pm 0.2 \degree$ and $\lambda_{\mathrm{A}} = -8.8 \pm 0.2 \degree$ respectively), yet still indicate slight misalignment. 

\begin{figure*}
\resizebox{\hsize}{!} 
        {\includegraphics[]{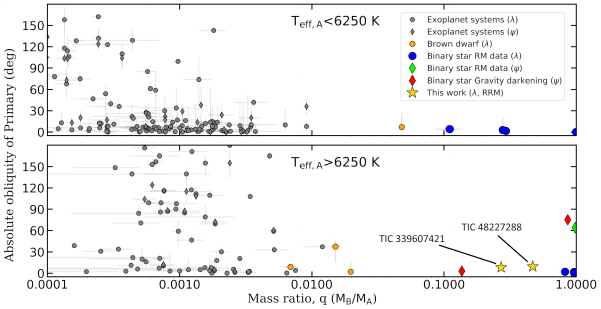}}
  \caption{Absolute values of the projected, $\lambda$, and true obliquities, $\psi$, of exoplanetary and binary star systems reported in the literature as a function of the companion to host mass ratio, $q$. Only primary star obliquities are shown for binary star systems. The two target systems that are the focus of this study and their corresponding RRM obliquities are shown as yellow stars.} 
     \label{fig:obliq vs q}
\end{figure*}

According to pre-main-sequence (PMS) evolutionary models (e.g. \citet{Baraffe_2015:1603}), both primary and secondary stars would have had significantly larger radii during their PMS phases. The current compact orbital configurations (a/$R_{\mathrm{A}} \approx$ 6.6 and 6.9 for TIC 48227288 and TIC 339607421, respectively) therefore suggest that the companions are unlikely to have formed in their current orbital positions, implying that a degree of orbital evolution has taken place. Potential mechanisms to drive that evolution include dynamical interactions with a third body, disk-driven migration, or perturbations from galactic tides. Analogous dynamical processes have been invoked for example by \citet{Mazeh_1979:1224} to explain close-in, misaligned hot Jupiter systems \citep[such as observed for example by][]{Addison_2013:450,Dorval_2020:522,Cabot_2021:1037}. In this respect TIC 48227288 and TIC 339607421 make for valuable case studies, as the low companion masses, small mass ratios, and compact orbits, place TIC 48227288 and TIC 339607421 in the transitional parameter space between classical binaries and hot Jupiter systems (see Figures \ref{fig:obliq vs q,M2} to \ref{fig:obliq vs q}).

\begin{figure*}
\resizebox{\hsize}{!} 
        {\includegraphics[]{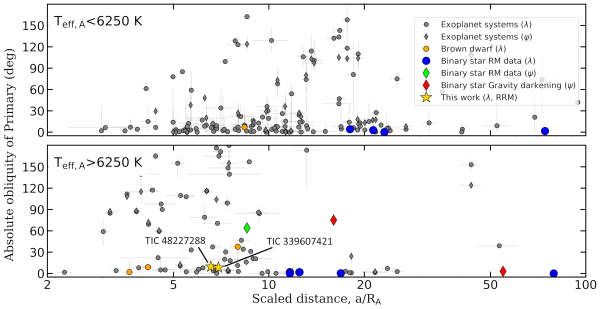}}
  \caption{Absolute values of the projected, $\lambda$, and true obliquities, $\psi$, of exoplanetary, brown dwarf and binary star systems reported in the literature as a function of the scaled distance, a/$R_\mathrm{A}$. Only primary star obliquities are shown for binary star systems. The two target systems that are the focus of this study and their corresponding RRM obliquities are shown as yellow stars. } 
     \label{fig:obliq vs a/R}
\end{figure*}

The present obliquities of these systems are shaped by both their dynamical histories and the effectiveness of tidal interactions in re-aligning the orbits. The strength of tidal interactions depends on the companion-to-host mass ratio, orbital separation, and the internal structure of the host star. Host stars on the main sequence with effective temperatures below the Kraft break, (here assumed to be \Teffprim{}=6250 K), have deeper convective envelopes that are more effective in dissipating the energy contained in tidal bulges generated by the nearby companion \citep{Zahn_1977:779,Albrecht_2012:1212}. Because of this, convective host stars are thought to be more effective in realigning orbits than host stars operating above the Kraft break which possess radiative envelopes. Among exoplanet systems, this dependence manifests as broader obliquity distributions at longer scaled separations and lower mass ratios for convective hosts (Figures \ref{fig:obliq vs a/R} and \ref{fig:obliq vs q}, top panels). These trends are not nearly so obvious in systems with radiative hosts (Figures \ref{fig:obliq vs a/R} and \ref{fig:obliq vs q}, bottom panels), indicating that tidal alignment is less efficient. For binary stars, obliquity trends are less clear due to limited data, though most misaligned systems appear to involve primaries above the Kraft break (again see Figures \ref{fig:obliq vs a/R} and \ref{fig:obliq vs q}).

\begin{figure*}
\resizebox{\hsize}{!} 
        {\includegraphics[]{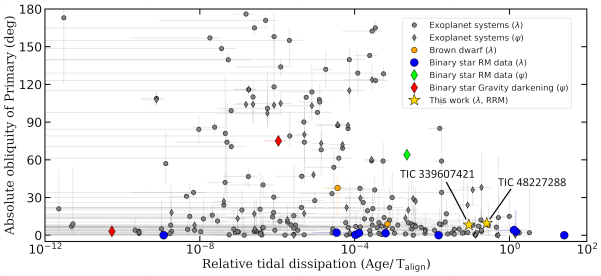}}
  \caption{Absolute values of the projected, $\lambda$, and true obliquities, $\psi$, of exoplanetary and binary star systems as a function of their relative alignment timescale. The two target systems that are the focus of this study and their corresponding RRM obliquities are shown as yellow stars.} 
     \label{fig:obliq vs age/T align}
\end{figure*}

System age is another key variable. Even in systems with modest tidal interaction, sufficient time can permit significant orbital realignment. To evaluate this, we adopt the tidal alignment timescale estimates from \citet{Albrecht_2012:326}, (Equations \ref{equat 2} and \ref{equat 3}), and compare these to system ages (Figure \ref{fig:obliq vs age/T align}). These models include a number of simplifications but offer a relative measure of tidal effectiveness between different systems. Among exoplanets, higher misalignment correlates with lower age-to-alignment-time ratios, however this pattern is again less evident for binary star systems. For our two target systems, the ages are comparable to the estimated alignment timescale (age/alignment time $\approx$ 0.25 and 0.1 for TIC 48227288 and TIC 339607421 respectively), suggesting that alignment processes have had considerable time to act and so it is possible that the extent of misalignment was historically greater that currently indicated by our modeling.


\begin{equation} \label{equat 2}
\frac{1}{\tau_{convective}} = \frac{1}{10\times10^9yr}q^2\left(\frac{a/R_{\star}}{40}\right)^{-6}
\end{equation}
\begin{equation} \label{equat 3}
\frac{1}{\tau_{radiative}} = \frac{1}{0.25\times5\times10^9yr}q^2\left(1+q\right)^{5/6}\left(\frac{a/R_{\star}}{6}\right)^{-17/2}
\end{equation}
\\

\begin{figure}
\resizebox{\hsize}{!} 
        {\includegraphics[]{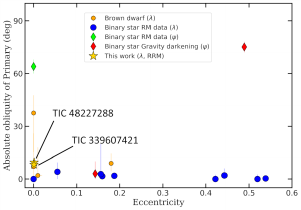}}
  \caption{Absolute values of the projected, $\lambda$, and true obliquities, $\psi$, of  binary star systems as a function of their orbital eccentricities. The two target systems that are the focus of this study and their corresponding RRM obliquities are shown as yellow stars.} 
     \label{fig:obliq vs ecc}
\end{figure}

Mechanisms such as Kozai–Lidov cycles \citep{Kozai_1962:307}, which can induce spin–orbit misalignment, also excite orbital eccentricity. Indeed, many binaries with measured obliquities exhibit significant eccentricities (Table \ref{table A1:previous studies}; Figure \ref{fig:obliq vs ecc}). However, both TIC 48227288B and TIC 339607421B follow circular orbits (e $\approx$ 0.001 in each case). Tidal theory predicts that orbital alignment proceeds more rapidly than orbital circularisation in such systems, since orbital angular momentum far exceeds stellar spin angular momentum \citep{Albrecht_2014:1214,Anderson_2017:1577,Lin_2017:1652}. Consequently, observing misalignment in a circular orbit, even a minor level of misalignment such as suggested by our modeling of the TIC 48227288 and TIC 339607421 systems, presents a theoretical challenge, as previously noted for the similar case of CV Velorum (\citet{Albrecht_2014:1214}, e=0, $\psi_A=67 ^{\circ}$). One explanation is ongoing dynamical perturbation by an unseen tertiary companion. As noted in \cite{Albrecht_2014:1214} this may indicate the continuing interaction with a third body that is producing spin-orbit misalignment despite the tidal damping taking place or, as pointed out by \citet{Justesen_2021:1252}, may hint at a more complicated formation migration history not well described by the current tidal theory.

\begin{figure}
\resizebox{\hsize}{!} 
        {\includegraphics[]{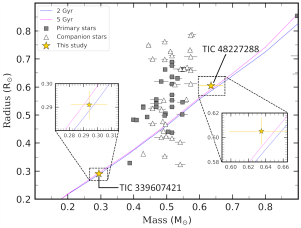}}
  \caption{Stellar radius as a function of stellar mass. The plot compares observed mass/radius for low mass stars against theoretical predictions. Theoretical mass-radius models \citep{Baraffe_2015:1603} for 2 Gyr and 5 Gyr models are shown as blue and red lines respectively. The two low mass companions from this study are highlighted as gold stars. Other data points are extracted from \citet{Garrido_2019:1619} and \citet{Cruz_2022:1618} where mass and radius uncertainties are <5$\%$ } 
     \label{fig:radius anomaly}
\end{figure}

Although not the primary objective of this study, our characterisation of the low-mass companions in our target binary systems contributes useful data with which to test the accuracy of low-mass stellar evolutionary models. Previous studies of short-period, (P~<~3~d), detached eclipsing binaries have reported that theoretical models systematically underestimate the radii of low-mass  (M < 0.7 \Msun{}) stars by as much as 5–20$\%$, with the discrepancy more pronounced for companion stars  \citep{Garrido_2019:1619,Cruz_2022:1618}. TIC 48227288B and TIC 339607421B both lie within this regime and with the modeling we undertook, we were able to achieve mass and radius estimates with <6\% and <2\% uncertainties, respectively for each companion. Figure \ref{fig:radius anomaly} compares the derived companion mass/radii with theoretical mass-radius model predictions, \citep{Baraffe_2015:1603}, for ages of 2 and 5 Gyr which bracket the estimated ages of our systems. Contrary to previous findings, we found that the theoretical models performed well in predicting the mass/radius characteristics of our two companions with TIC 339607421 falling within the 2 Gyr and 5~Gyr model predictions and TIC 48227288B agreeing within stated uncertainties.


\section{Conclusions} \label{section:conclusions}
In this study we present an analysis of \textit{TESS} photometric data and \minaus{} radial velocity data to characterize two single-lined spectroscopic binary star (SB1) systems and estimate the obliquities of their primary stars. 

TIC 339607421 comprises an F6V primary ($M_A~=~1.09~\pm~0.04$~\Msun{}, $R_A = 1.21_{-0.02}^{+0.03}$~\Rsun{}) orbited by an M-dwarf companion ($M_B=0.294 \pm {+0.013}$~\Msun{}, $R_B=0.291 \pm 0.006$~\Rsun{}) in a short period ($P_B\sim$2.4 d), circular ($e\sim$0.001) orbit. TIC 48227288 consists of an F3V primary ($M_A = 1.36^{0.06}_{-0.08}$~\Msun{}, $R_A = 1.61 \pm 0.03$~\Rsun{}) orbited by a late K-class companion ($M_B=0.635 \pm {0.037}$ \Msun{}, $R_B=0.605 \pm 0.011$ \Rsun{}) in a similarly short-period ($P_B\sim$2.9 d), near circular ($e\sim$~0.001) orbit.

Analysis of the anomalous RV perturbation using \allesfitter{} yielded marginally misaligned prograde orbits for both systems, (TIC 48227288: $\lambda_{\mathrm{A}} = -17.8^{+1.9}_{-2.0} \degree$ , TIC 339607421: $\lambda_{\mathrm{A}} = -14.7^{+5.4}_{-5.9} \degree$). A second RRM analysis yielded slightly lower obliquities, (TIC 48227288: $\lambda_{\mathrm{A}} = -9.5 \pm 0.2 \degree$ and TIC 339607421: $\lambda_{\mathrm{A}} = -8.8 \pm 0.2 \degree$) but confirmed the minor misalignment inferred from the classical RV analysis.

These results contribute to the limited sample of binary systems involving cool primaries with measured obliquities. Notably, TIC 339607421 and TIC 48227288 represent the shortest period systems within this sample.  The detection of slight misalignment for companions in circular orbits, along with a similar findings for systems such as CV Vel, suggests that current models of binary formation and orbital evolution may be incomplete and that further investigation of similar systems are warranted to enable a better understanding of the formation and evolution of binary star systems.

\section*{Acknowledgements}

We respectfully acknowledge the traditional custodians of all lands throughout Australia, and recognize their continued cultural and spiritual connection to the land, waterways, cosmos, and community. We pay our deepest respects to all Elders, ancestors and descendants of the Giabal, Jarowair, and Kambuwal nations, upon whose lands the Minerva-Australis facility at Mt Kent is situated.

We also wish to thank the anonymous referee for their many suggestions and questions about this work. Their advice has led to significant improvements.

\minaus{}-Australis is supported by Australian Research Council LIEF Grant LE160100001, Discovery Grants DP180100972 and DP220100365, Mount Cuba Astronomical Foundation, and institutional partners University of Southern Queensland, UNSW Sydney, MIT, Nanjing University, George Mason University, University of Louisville, University of California Riverside, University of Florida, and The University of Texas at Austin.
Funding for the \textit{TESS} mission is provided by NASA’s Science Mission directorate. We acknowledge the use of public \textit{TESS} Alert data from pipelines at the \textit{TESS} Science Office and at the \textit{TESS} Science Processing Operations Center. This research has made use of the Exoplanet Follow-up Observation Program website, which is operated by the California Institute of Technology, under contract with the National Aeronautics and Space Administration under the Exoplanet Exploration Program. This paper includes data collected by the \textit{TESS} mission, which are publicly available from the Mikulski Archive for Space Telescopes (MAST).\\

TW is supported by the Australian Government Research Training Program (RTP) Scholarship.

\textbf{Facilities}: \minaus{}-Australis. The UniSQ High Performance Computing (HPC) facility, Fawkes.

\textbf{Software}: Astropy \citep{astropy:2022}, Matplotlib \citep{Matplotlib_Hunter:2007}, Allesfitter \citep{Gunther_2021:1318}, Radvel \citep{Fulton_2018:274}, AstroAriadne \citep{Vines_2022:1420}, iSpec \citep{blanco_2014:1396}

\section*{Data Availability}

The \minaus{} radial velocity data and \textit{TESS} photometric data underlying this article are available in full in the online supporting material. The \textit{TESS} data used in this paper are also available via NASA's Mikulski Archive for Space telescopes:
\href{https://mast.stsci.edu/portal/Mashup/Clients/Mast/Portal.html}{https://mast.stsci.edu/portal/Mashup/Clients/Mast/Portal.html}


\clearpage
\bibliographystyle{mnras}
\bibliography{refs} 



\onecolumn
\appendix
\section{ Previous studies} \label{Appendix_A}

\setcounter{table}{0}
\renewcommand{\thetable}{A\arabic{table}}


\begin{small}
\renewcommand*{\arraystretch}{1.4}

\begin{longtable}{lcccccccccc}

\caption{\label{table A1:previous studies} Previous obliquity studies involving Binary star systems}\\
\hline\hline
 &  & Period &$R_A,R_B$ &  &  \multicolumn{4}{c}{Obliquity, deg} & Method$\vdag$ & Reference$\vddag$\\  
\cline{6-9}
System & Spectral type & [d] &  [\Rsun] & Ecc. &$\lambda_A$&$\lambda_B$ &$\psi_A$ &$\psi_B$& &   \\  

\hline
\endfirsthead
\caption{Previous obliquity studies involving Binary star systems (continued).}\\
\hline\hline
&  & Period &$R_A,R_B$ &  &  \multicolumn{4}{c}{Obliquity, deg} & Method$\vdag$ & Reference$\vddag$\\  
\cline{6-9}
System & Spectral type & [d] &  [\Rsun] & Ecc. &$\lambda_A$&$\lambda_B$ &$\psi_A$ &$\psi_B$& &   \\  
\hline
\endhead
\hline
\endfoot


$\beta$ Lyrae      & Be+B6-8II      & 12.9   & 6.0,15.2  & <0.01      & \multicolumn{4}{c}{aligned?} & 1          & 1  \\
V1010 Oph          & A7IV-V+        & 0.7    & 2.0,1.3   & 0.072      & \multicolumn{4}{c}{aligned?} & 1          & 2  \\
W Umi              & A3V+G9IV       & 1.7    & 3.6,3.1   & $\equiv 0$ & \multicolumn{4}{c}{aligned?} & 1          & 3  \\
$\delta$ Lib       & A0V+K0IV       & 2.3    & 4.1,4.2   & 0.069      & \multicolumn{4}{c}{aligned?} & 1          & 4  \\
V505 Sgr (a)       & A2V+GIV        & 1.2    & 2.1,2.4   & $\equiv 0$ & \multicolumn{4}{c}{aligned?} & 1          & 5  \\
V505 Sgr (b)       & A2V+GIV        & 1.2    & 2.1,2.4   & $\equiv 0$ & \multicolumn{4}{c}{aligned?} & 1          & 6  \\
AI Dra             & A0V+F9.5V      & 1.2    & 2.1,2.4   & $\equiv 0$ & \multicolumn{4}{c}{aligned?} & 1          & 5  \\
X Tri              & A3V+G3IV       & 1.0    & 1.7,2.0   & $\equiv 0$ & \multicolumn{4}{c}{aligned?} & 1          & 3  \\
RZ Cas             & A3V+KIV        & 1.2    & 1.6,1.9   & $\equiv 0$ & \multicolumn{4}{c}{aligned?} & 1          & 3  \\
U Sge              & B8.5V+G3III    & 3.4    & 4.0,5.9   & 0.04       & \multicolumn{4}{c}{aligned?} & 1          & 3  \\
WW Cyg             & B8V+G4III      & 3.3    & 4.2,5.7   & $\equiv 0$ & \multicolumn{4}{c}{aligned?} & 1          & 3  \\
Y Leo              & A3V+K3IV       & 1.7    & 1.9,2.5   & $\equiv 0$ & \multicolumn{4}{c}{aligned?} & 1          & 3  \\
RX Hya             & A8+K0IV        & 2.3    & 1.8,2.6   & $\equiv 0$ & \multicolumn{4}{c}{aligned?} & 1          & 3  \\
$\beta$ Per (a)    & B8+K2IV        & 2.9    & 2.7,3.5   & $\equiv 0$ & \multicolumn{4}{c}{aligned?} & 1          & 7  \\
$\beta$ Per (b)    & B8+K2IV        & 2.9    & 2.7,3.5   & $\equiv 0$ & \multicolumn{4}{c}{aligned?} & 1          & 8  \\
RW Gem             & B5-B6V+F0III   & 2.9    & 3.4,4.5   & $\equiv 0$ & \multicolumn{4}{c}{aligned?} & 1          & 3  \\
Y Psc              & A3V+K2IV       & 3.8    & 2.7,3.7   & 0.12       & \multicolumn{4}{c}{aligned?} & 1          & 3  \\
TV Cas             & A2V+G1IV       & 1.8    & 3.2,3.3   & $\equiv 0$ & \multicolumn{4}{c}{aligned?} & 1          & 3  \\
ST Per             & A3V+KIV        & 2.6    & 2.3,3.0   & $\equiv 0$ & \multicolumn{4}{c}{aligned?} & 1          & 3  \\
U Cep              & B7V+G8III      & 2.5    & 2.8,5.2   & $\equiv 0$ & \multicolumn{4}{c}{aligned?} & 1          & 3  \\
TX Uma             & B8V+F7-F8III   & 3.1    & 2.7,4.1   & $\equiv 0$ & \multicolumn{4}{c}{aligned?} & 1          & 3  \\
W Del              & A0-B9.5V+K0IV  & 4.8    & 2.4,4.6   & 0.2        & \multicolumn{4}{c}{aligned?} & 1          & 3  \\
DE Dra             & B0V+B2V        & 5.3    & 2.9,1.1   & 0.02       & \multicolumn{4}{c}{misaligned?} & 1       & 9  \\
SW Cyg             & A2V+K1IV       & 4.6    & 2.6,4.3   & 0.3        & \multicolumn{4}{c}{aligned?} & 1          & 3  \\
RY Per             & B4V+F0III      & 6.9    & 4.1,8.1   & 0.21       & \multicolumn{4}{c}{aligned?} & 1          & 3  \\
RZ Sct             & B2II+A0II-III  & 15.2   & 15.8,15.9 & $\equiv 0$ & \multicolumn{4}{c}{aligned?} & 1          & 3  \\
AQ Peg             & A2V+K1IV       & 5.6    & 2.7,4.8   & 0.24       & \multicolumn{4}{c}{aligned?} & 1          & 3  \\
RY Gem             & A2V+K0III-K1IV & 9.3    & 2.4,6.2   & 0.16       & \multicolumn{4}{c}{aligned?} & 1          & 3  \\
V1143 Cyg (a)      & F5V+F5V        & 7.6    & 1.3,1.3   & 0.54       & \multicolumn{4}{c}{aligned?} & 1          & 10  \\
DI Her (a)         & B5V+B5V        & 10.6   & 2.7,2.5   & 0.49       & \multicolumn{4}{c}{aligned?} & 1          & 11  \\
NY Ceph (a)        & B0V+B2V        & 15.3   & 6.0,5.8   & 0.44       & \multicolumn{4}{c}{aligned?} & 1          & 12  \\
NY Ceph (b)        & B0V+B2V        & 15.3   & 6.0,5.8   & 0.44       & $-2.0\pm{4.0}$          &                         &                         &                         & 2          & 13  \\
DI Her (b)         & B5V+B5V        & 10.6   & 2.7,2.5   & 0.49       & $-72.0\pm{4.0}$         & $84.0\pm{8.0}$          &                         &                         & 2          & 14  \\
DI Her (c)         & B5V+B5V        & 10.6   & 2.7,2.5   & 0.49       & $-74.0_{-3.0}^{+2.0}$   & $79.0_{-3.0}^{+2.0}$    & $75.0\pm{3.0}$          & $80.0\pm{3.0}$          & 3          & 15  \\
V1143 Cyg (b)      & F5V+F5V        & 7.6    & 1.3,1.3   & 0.54       & $-0.3\pm{1.5}$          & $1.2\pm{1.6}$           &                         &                         & 2          & 16  \\
EP Cru (a)         & B5V+B5V        & 11.1   & 3.6,3.5   & 0.19       & $1.8\pm{1.6}$           & <17                     &                         &                         & 2          & 17  \\
EP Cru (b)         & B5V+B5V        & 11.1   & 3.6,3.5   & 0.19       &                         &                         & $6.0_{-4.5}^{+7.1}$     &                         & 4          & 18  \\
Kepler 16 (a)      & K7+            & 41.1   & 0.6,0.2   & 0.16       & $-1.6\pm{2.4}$          &                         & <18.3                   &                         & 2          & 19  \\
Kepler 16 (b)      & K7+            & 41.1   & 0.6,0.2   & 0.16       &                         &                         & $8.4_{-5.2}^{+10.2}$    &                         & 4          & 18  \\
KOI-368.01 (a)     & A+M            & 110.3  & 2.3,0.2   & 0.14       & $36.0_{-17.0}^{+23.0}$  &                         & $69.0_{-10.0}^{+9.0}$   &                         & 3          & 20  \\
KOI-368.01 (b)     & A+M            & 110.3  & 2.3,0.2   & 0.14       & $10.0\pm{2.0}$          &                         & $3.0\pm{7.0}$           &                         & 3          & 21  \\
KOI-368.01 (c)     & A+M            & 110.3  & 2.3,0.2   & 0.14       &                         &                         & $12.9_{-4.8}^{+6.2}$    &                         & 4          & 18  \\
AI Hya             & F2m+F0V        & 8.3    & 3.9,2.8   & 0.23       &                         &                         & $14.1_{-7.4}^{+11.1}$   &                         & 4          & 18  \\
Y Cyg              & O9.5 IV+O9.5IV & 3.0    & 5.8,5.8   & 0.15       &                         &                         & $21.7_{-12.2}^{+11.2}$  &                         & 4          & 18  \\
QX car             & B2V+B2V        & 4.5    & 4.3,4.1   & 0.28       &                         &                         & $22.6_{-12.0}^{+12.2}$  &                         & 4          & 18  \\
V541 Cyg           & B9.5+B9.5      & 15.3   & 1.9,1.8   & 0.47       &                         &                         & $15.9_{-8.5}^{+21.7}$   &                         & 4          & 18  \\
FM Leo (a)         & F6V+F6V        & 6.7    & 1.8,1.2   & 0          & $0.0\pm{1.1}$           & $0.3\pm{27.4}$          &                         &                         & 5          & 22  \\
FM Leo (b)         & F6V+F6V        & 6.7    & 1.8,1.2   & 0          &                         &                         & $8.1_{-7.2}^{+36.2}$    &                         & 4          & 18  \\
NN Del (a)         & F8+            & 99.3   & 1.6,2.2   & 0.52       & $0.0\pm{2.6}$           &                         &                         &                         & 5          & 22  \\
NN Del (b)         & F8+            & 99.3   & 1.6,2.2   & 0.52       &                         &                         & $8.1_{-5.6}^{+37.5}$    &                         & 4          & 18  \\
V963 Cen (a)       & G2V+G2V        & 15.3   & 1.4,1.4   & 0.42       & $0.0\pm{2.4}$           &                         &                         &                         & 5          & 22  \\
V963 Cen (b)       & G2V+G2V        & 15.3   & 1.4,1.4   & 0.42       &                         &                         & $8.1_{-5.7}^{+37.5}$    &                         & 4          & 18  \\
GG Lup             & B7V+B9V        & 1.8    & 2.4,1.8   & 0.15       &                         &                         & $30.4_{-16.9}^{+17.6}$  &                         & 4          & 18  \\
V364 Lac           & A4m+A3m        & 7.4    & 3.3,3.0   & 0.29       &                         &                         & $30.4_{-17.9}^{+19.4}$  &                         & 4          & 18  \\
TOI 1338 (a)       & F8+            & 14.6   & 1.3,0.3   & 0.16       & $-2.8\pm{17.1}$         &                         &                         &                         & 6          & 23  \\
TOI 1338 (b)       & F8+            & 14.6   & 1.3,0.3   & 0.16       &                         &                         & $25.6_{-15.0}^{+26.3}$  &                         & 4          & 18  \\
V530 Ori           & G1V+M1V        & 6.1    & 1.0,0.6   & 0.09       &                         &                         & $37.0_{-18.8}^{+16.1}$  &                         & 4          & 18  \\
V459 Cas           & A1m+A1m        & 8.5    & 2.0,2.0   & 0.02       &                         &                         & $32.5_{-18.0}^{+19.8}$  &                         & 4          & 18  \\
V578 Mon           & B1V+B2V        & 2.4    & 5.4,4.3   & 0.08       &                         &                         & $38.2_{-19.6}^{+15.6}$  &                         & 4          & 18  \\
PV Cas             & B9.5V+B9.5V    & 1.8    & 2.3,2.3   & 0.03       &                         &                         & $39.7_{-20.2}^{+16.7}$  &                         & 4          & 18  \\
V478 Cyg           & 09.5V+09.5V    & 2.9    & 7.3,7.2   & 0.02       &                         &                         & $41.5_{-22.9}^{+17.5}$  &                         & 4          & 18  \\
PT vel             & A1V+A6V        & 1.8    & 2.1,1.6   & 0.13       &                         &                         & $40.6_{-22.5}^{+22.0}$  &                         & 4          & 18  \\
V636 Cen           & G0V+K2V        & 4.3    & 1.0,0.8   & 0.13       &                         &                         & $46.4_{-24.0}^{+20.0}$  &                         & 4          & 18  \\
IQ Per             & B8V+A6V        & 1.7    & 2.4,1.5   & 0.07       &                         &                         & $44.2_{-23.4}^{+23.0}$  &                         & 4          & 18  \\
IT Cas             & F3V+F3V        & 3.9    & 1.6,1.6   & 0.09       &                         &                         & $43.3_{-22.4}^{+24.4}$  &                         & 4          & 18  \\
EK Cep             & A1V+G5V        & 4.4    & 1.6,1.3   & 0.11       &                         &                         & $46.7_{-23.9}^{+22.4}$  &                         & 4          & 18  \\
J1219-39b          & K0V+           & 6.8    & 0.8,0.1   & 0.06       & $-4.1_{-5.3}^{+4.8}$    &                         &                         &                         & 5          & 24  \\
CV Vel             & B2.5V+B2.5V    & 6.9    & 4.1,4.0   & 0          & $52.0\pm{6.0}$          & $-3.7\pm{7.0}$          & $64.0\pm{4.0}$          & $46.0\pm{9.0}$          & 2          & 25  \\
AS Cam             & B8V+B9.5V      & 3.4    & 2.6,2.0   & 0.16       &                         &                         & $87.9_{-2.5}^{+2.1}$    &                         & 4          & 18  \\
AI Phe             & K0IV+F7V       & 24.6   & 2.9,1.8   & 0.19       &                         & $-87.0\pm{17.0}$        &                         &                         & 5          & 22  \\

\end{longtable}
\end{small}
\textbf{$\vdag$ Methods:} (1) RM effect - visual assessment from plot of RV perturbation; (2) RM effect - velocity profile modelling; (3) Gravity darkening; (4) Analysis of apsidal motion; (5) RM effect - broadening function (6) RM effect - reloaded RM \\

\textbf{$\vddag$ References:} (1) \citet{Rossiter_1924:260}; (2) \citet{Worek_1988:1241};  (3) \citet{Twigg_1979:1439}; (4) \citet{Bakis_2006:1242}; (5) \citet{Worek_1996:1243}; (6) \citet{Lazaro_2006:1428}; (7) \citet{McLaughlin_1924:228}; (8) \citet{Struve_1931:1265}; (9)\citet{Hube_1982:1245}; (10) \citet{Andersen_1987:1445}; (11) \citet{Popper_1982:1444}; (12) \citet{Holmgren_1990:1443}; (13) \citet{Albrecht_2011:1201}; (14) \citet{Albrecht_2009:1200}; (15) \citet{Liang_2022:1249}; (16) \citet{Albrecht_2007:1216}; (17) \citet{Albrecht_2013:1215}; (18) \citet{Marcussen_2022:1213}; (19) \citet{Winn_2011:1203}; (20) \citet{Zhou_2013:1239}; (21) \citet{Ahlers_2014:1251}; (22) \citet{Sybilski_2018:1247}; (23) \citet{Kunovac_2020:1209};  (24) \citet{Triaud_2013:434}; (25) \citet{Albrecht_2014:1214}

\renewcommand{\arraystretch}{1.5}   

\newpage
\renewcommand{\thetable}{B\arabic{table}}
\section{ Additional fitted and derived data} \label{Appendix_B}


\begin{longtable}{lccc}
\caption{\label{table:additional TIC 48227288 fit results} Median and 68\% confidence intervals of additional astrophysical parameters derived for TIC 48227288 by \allesfitter{}. Priors are shown as uniform \uni(a,b) or normal \norm$(\mu,\sigma)$. Notes: $\vdag$ preferred solution}\\
\hline\hline

Parameter & Prior & Best Fit$\vdag{}$ & Best Fit (flattened light curve) \\  

\hline
\endfirsthead
\caption{continued.}\\
\hline\hline
Parameter & Prior & Best Fit & Best Fit (flattened light curve)$\vdag{}$ \\  

\hline
\endhead
\hline
\endfoot

\textbf{Flux error scaling (ln relative flux)} & & & \\
$\ln{\sigma_{F,(120s)}}$ &                    \uni$(-10.0,-3.0)$ &   $-6.265\pm0.003$ &                    $-6.606\pm0.003$\\

\textbf{Transformed limb darkening} & & & \\
$q1_{TESS,A}$ &                               \uni$(0,1)$ &          $0.30_{-0.09}^{+0.06}$ &              $0.36_{-0.05}^{+0.04}$\\
$q2_{TESS,A}$ &                               \uni$(0,1)$ &          $0.15_{-0.10}^{+0.21}$ &              $0.09_{-0.06}^{+0.07}$\\
$q1_{TESS,B}$ &                               \uni$(0,1)$ &          $0.48_{-0.14}^{+0.19}$ &              $0.28_{-0.07}^{+0.11}$\\
$q2_{TESS,B}$ &                               \uni$(0,1)$ &          $0.64_{-0.24}^{+0.22}$ &              $0.77_{-0.24}^{+0.16}$\\

\textbf{RV jitter  (ln \kms)} & & & \\
$\ln{\sigma_\mathrm{jitter;T1(1800s)}}$ &     \uni$(-20,3)$ &        $-8.09_{-5.99}^{+6.29}$ &             $-9.92_{-6.19}^{+6.47}$\\
$\ln{\sigma_\mathrm{jitter;T3 (1500s)}}$ &    \uni$(-20,3)$ &        $-9.81_{-5.81}^{+4.48}$ &             $-11.86_{-5.30}^{+5.87}$\\
$\ln{\sigma_\mathrm{jitter;T3RM (1500s)}}$ &  \uni$(-20,3)$ &        $-5.94_{-6.88}^{+4.15}$ &             $-8.48_{-7.08}^{+6.53}$\\
$\ln{\sigma_\mathrm{jitter;T3 (1800s)}}$ &    \uni$(-20,3)$ &        $-8.05_{-6.44}^{+5.26}$ &             $-10.34_{-5.95}^{+6.43}$\\
$\ln{\sigma_\mathrm{jitter;T4RM (1500s)}}$ &  \uni$(-20,3)$ &        $-13.33_{-4.45}^{+5.83}$ &            $-5.89_{-4.12}^{+3.28}$\\
$\ln{\sigma_\mathrm{jitter;T4 (1500s)}}$ &    \uni$(-20,3)$ &        $-7.97_{-4.94}^{+3.92}$ &             $-10.62_{-5.95}^{+5.37}$\\
$\ln{\sigma_\mathrm{jitter;T4 (1800s)}}$ &    \uni$(-20,3)$ &        $-7.28_{-5.51}^{+4.76}$ &             $-10.73_{-6.06}^{+6.75}$\\
$\ln{\sigma_\mathrm{jitter;T5 (1500s)}}$ &    \uni$(-20,3)$ &        $-5.92_{-5.73}^{+5.07}$ &             $-8.74_{-6.32}^{+6.56}$\\

\textbf{RV baseline } & & &  \\
$\mathrm{GP\ amplitude} \ln{\mathrm{a;T1}}$ & \uni$(-20,20)$ &       $0.73_{-0.84}^{+1.02}$ &              $0.61_{-0.67}^{+0.84}$\\
$\mathrm{GP\ time\ scale} \ln{\mathrm{c;T1}}$ & \uni$(-20,20)$ &       $5.16_{-7.43}^{+8.97}$ &              $3.49_{-6.10}^{+8.08}$\\
$\mathrm{GP\ amplitude}\ln{\mathrm{a;T3}}$ &  \uni$(-20,20)$ &       $0.55_{-0.34}^{+0.47}$ &              $0.57_{-0.33}^{+0.47}$\\
$\mathrm{GP\ time\ scale} \ln{\mathrm{c;T3}}$ & \uni$(-20,20)$ &       $-2.62_{-0.60}^{+0.47}$ &             $-2.44_{-0.63}^{+0.52}$\\
$\mathrm{GP\ amplitude}\ln{\mathrm{a;T3RM}}$ & \uni$(-20,20)$ &       $-4.09_{-6.13}^{+1.02}$ &             $-3.79_{-2.76}^{+0.83}$\\
$\mathrm{GP\ time\ scale} \ln{\mathrm{c;T3RM}}$ & \uni$(-20,20)$ &       $7.36_{-8.42}^{+7.78}$ &              $8.32_{-5.84}^{+7.36}$\\
$\mathrm{GP\ amplitude}\ln{\mathrm{a;T4}}$ &  \uni$(-20,20)$ &       $0.85_{-0.40}^{+0.58}$ &              $0.77_{-0.39}^{+0.50}$\\
$\mathrm{GP\ time\ scale} \ln{\mathrm{c;T4}}$ & \uni$(-20,20)$ &       $-3.12_{-0.64}^{+0.51}$ &             $-3.08_{-0.58}^{+0.49}$\\
$\mathrm{GP\ amplitude}\ln{\mathrm{a;T4RM}}$ & \uni$(-20,20)$ &       $-10.40_{-5.68}^{+5.14}$ &            $-6.90_{-5.23}^{+2.89}$\\
$\mathrm{GP\ time\ scale} \ln{\mathrm{c;T4RM}}$ & \uni$(-20,20)$ &       $7.99_{-11.77}^{+8.15}$ &             $3.28_{-14.44}^{+10.53}$\\
$\mathrm{GP\ amplitude}\ln{\mathrm{a;T5}}$ &  \uni$(-20,20)$ &       $1.86_{-1.10}^{+1.83}$ &              $1.77_{-0.97}^{+1.74}$\\
$\mathrm{GP\ time\ scale}\ln{\mathrm{c;T5}}$ & \uni$(-20,20)$ &       $-4.93_{-2.32}^{+1.32}$ &             $-4.84_{-2.43}^{+1.28}$\\

\hline

\end{longtable}

\newpage


\begin{longtable}{lccc}
\caption{\label{table:TIC 339607421 additional fit results} Median and 68\% confidence intervals of additional astrophysical parameters derived for TIC 339607421 by \allesfitter{}. Priors are shown as uniform \uni(a,b) or normal \norm$(\mu,\sigma)$.Notes:$\vdag{}$ preferred solution.}\\
\hline\hline

Parameter & Prior & Best Fit & Best Fit (flattened light curve)$\vdag{}$ \\  

\hline
\endfirsthead
\caption{continued.}\\
\hline\hline
Parameter & Prior & Best Fit & Best Fit (flattened light curve)$\vdag{}$ \\  

\hline
\endhead
\hline
\endfoot

\textbf{Flux error scaling (ln relative flux)} & & & \\
$\ln{\sigma_{F,(120s)}}$ &                    \uni$(-10.0,-3.0)$ &   $-6.506\pm0.003$ &                    $-7.546\pm0.003$\\
$\ln{\sigma_{F,(600s)}}$ &                    \uni$(-10.0,-3.0)$ &   $-5.818\pm0.008$ &                    $-7.987\pm0.008$\\

\textbf{Transformed limb darkening} & & & \\
$q1_{TESS,A}$ &                               \uni$(0,1)$ &          $0.16_{-0.03}^{+0.04}$ &              $0.10\pm0.01$\\
$q2_{TESS,A}$ &                               \uni$(0,1)$ &          $0.36_{-0.22}^{+0.27}$ &              $0.47_{-0.18}^{+0.15}$\\
$q1_{TESS,B}$ &                               \uni$(0,1)$ &          $0.44_{-0.27}^{+0.31}$ &              $0.73_{-0.18}^{+0.17}$\\
$q2_{TESS,B}$ &                               \uni$(0,1)$ &          $0.41_{-0.26}^{+0.33}$ &              $0.56_{-0.19}^{+0.22}$\\

\textbf{RV jitter  (ln \kms)} & & & \\
$\ln{\sigma_\mathrm{jitter;T1(1800s)}}$ &     \uni$(-20,3)$ &        $-0.19\pm0.27$ &                      $-0.24_{-0.25}^{+0.27}$\\
$\ln{\sigma_\mathrm{jitter;T3RM (900s)}}$ &   \uni$(-20,3)$ &        $-6.06_{-8.55}^{+4.41}$ &             $-5.78_{-8.18}^{+4.13}$\\
$\ln{\sigma_\mathrm{jitter;T3 (900s)}}$ &     \uni$(-20,3)$ &        $-10.28_{-6.29}^{+5.56}$ &            $-9.85_{-6.06}^{+5.21}$\\
$\ln{\sigma_\mathrm{jitter;T3 (1800s)}}$ &    \uni$(-20,3)$ &        $-0.20_{-0.20}^{+0.22}$ &             $-0.25_{-0.20}^{+0.22}$\\
$\ln{\sigma_\mathrm{jitter;T4RM (900s)}}$ &   \uni$(-20,3)$ &        $-1.19_{-0.28}^{+0.24}$ &             $-1.19_{-0.30}^{+0.25}$\\
$\ln{\sigma_\mathrm{jitter;T4 (900s)}}$ &     \uni$(-20,3)$ &        $-10.91_{-5.82}^{+5.86}$ &            $-11.52_{-5.53}^{+6.10}$\\
$\ln{\sigma_\mathrm{jitter;T4 (1800s)}}$ &    \uni$(-20,3)$ &        $-0.13_{-0.18}^{+0.19}$ &             $-0.17_{-0.17}^{+0.19}$\\
$\ln{\sigma_\mathrm{jitter;T5RM (900s)}}$ &   \uni$(-20,3)$ &        $-10.07_{-6.70}^{+7.06}$ &            $-4.78_{-5.27}^{+3.17}$\\
$\ln{\sigma_\mathrm{jitter;T5 (1800s)}}$ &    \uni$(-20,3)$ &        $-0.58\pm0.33$ &                      $-0.62_{-0.33}^{+0.32}$\\

\textbf{RV offset  (\kms)} & & &  \\
$\Delta \mathrm{RV}_\mathrm{T1 (1800s)}$ &    \uni$(6,26)$ &         $16.06_{-0.29}^{+0.28}$ &             $16.09\pm0.28$\\
$\Delta \mathrm{RV}_\mathrm{T3RM (900s)}$ &   \uni$(6,26)$ &         $16.24\pm0.08$ &                      $16.22\pm0.09$\\
$\Delta \mathrm{RV}_\mathrm{T3 (900s)}$ &     \uni$(6,26)$ &         $16.43\pm0.07$ &                      $16.39\pm0.07$\\
$\Delta \mathrm{RV}_\mathrm{T3 (1800s)}$ &    \uni$(6,26)$ &         $15.83_{-0.22}^{+0.23}$ &             $15.83_{-0.22}^{+0.21}$\\
$\Delta \mathrm{RV}_\mathrm{T4RM (900s)}$ &   \uni$(6,26)$ &         $15.94\pm0.10$ &                      $15.92\pm0.10$\\
$\Delta \mathrm{RV}_\mathrm{T4 (900s)}$ &     \uni$(6,26)$ &         $16.31\pm0.07$ &                      $16.29\pm0.07$\\
$\Delta \mathrm{RV}_\mathrm{T4 (1800s)}$ &    \uni$(6,26)$ &         $15.75_{-0.19}^{+0.20}$ &             $15.72\pm0.20$\\
$\Delta \mathrm{RV}_\mathrm{T5RM (900s)}$ &   \uni$(6,26)$ &         $16.02_{-0.12}^{+0.11}$ &             $16.00_{-0.11}^{+0.12}$\\
$\Delta \mathrm{RV}_\mathrm{T5 (1800s)}$ &    \uni$(6,26)$ &         $15.92_{-0.20}^{+0.21}$ &             $15.92_{-0.19}^{+0.20}$\\

\hline

\end{longtable}

\bsp	
\label{lastpage}
\end{document}